\documentclass{aa}  

\usepackage{graphicx}
\usepackage[varg]{txfonts}
\usepackage{graphicx}	
\usepackage{amsmath}	
\usepackage{xcolor}
\usepackage{tablefootnote}
\usepackage[T1]{fontenc}
\usepackage{booktabs,siunitx}
\newcommand{\orcid}[1]{\href{https://orcid.org/#1}{\includegraphics[width=9pt]{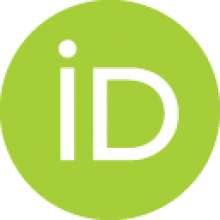}}}
\usepackage{natbib,twoopt}
\usepackage[breaklinks=true]{hyperref}[links=true, link={blue}, cite={blue}]
\hypersetup{
    colorlinks = true,
    linkcolor = {blue},
    citecolor = {blue}
}

\bibpunct{(}{)}{;}{a}{}{,} 
\makeatletter
\newcommandtwoopt{\citeads}[3][][]{\href{http://adsabs.harvard.edu/abs/#3}%
{\def\hyper@linkstart##1##2{}%
\let\hyper@linkend\@empty\citealp[#1][#2]{#3}}}
\newcommandtwoopt{\citepads}[3][][]{\href{http://adsabs.harvard.edu/abs/#3}%
{\def\hyper@linkstart##1##2{}%
\let\hyper@linkend\@empty\citep[#1][#2]{#3}}}
\newcommandtwoopt{\citetads}[3][][]{\href{http://adsabs.harvard.edu/abs/#3}%
{\def\hyper@linkstart##1##2{}%
\let\hyper@linkend\@empty\citet[#1][#2]{#3}}}
\newcommandtwoopt{\citeyearads}[3][][]%
{\href{http://adsabs.harvard.edu/abs/#3}
{\def\hyper@linkstart##1##2{}%
\let\hyper@linkend\@empty\citeyear[#1][#2]{#3}}}
\makeatother

\begin{document} 

\titlerunning{The role of AGN feedback in dwarf galaxies}
\authorrunning{Arjona-Gálvez E., Di Cintio A. \& Grand R.}

\title{The role of AGN feedback on the evolution of dwarf galaxies from cosmological simulations:}

\subtitle{SMBHs suppress star formation in low-mass galaxies}

\author{Elena Arjona-Gálvez \orcid{0000-0002-0462-7519} \inst{1,2}, Arianna Di Cintio \orcid{0000-0002-9856-1943} \inst{2,1}\and Robert J. J. Grand \orcid{0000-0001-9667-1340} \inst{3}}

\institute{Instituto de Astrofísica de Canarias, Calle Via Láctea s/n, E-38206 La Laguna, Tenerife, Spain\\
\email{eag@iac.es}
\and 
Universidad de La Laguna, Avda. Astrofísico Fco. Sánchez, E-38205 La Laguna, Tenerife, Spain\\
\email{alu0101295794@ull.edu.es}
\and
Astrophysics Research Institute, Liverpool John Moores University, 146 Brownlow Hill, Liverpool, L3 5RF, U}

\date{Received XXX, XXXX; accepted YYY, YYY}

 
  \abstract
   {}
   {Recent observational studies suggest that feedback from active galactic nuclei (AGNs) may play an important role in the formation and evolution of low-mass dwarf galaxies, an issue that has received little attention from a theoretical perspective.}
   {We investigate this using two sets of 12 cosmological magneto-hydrodynamical simulations of the formation of dwarf galaxies: one set using a version of the \texttt{AURIGA} galaxy formation physics model including AGN feedback and a parallel set with AGN feedback turned off.}
   {We show that the full-physics AGN runs satisfactorily reproduce several scaling relations, including the M$_{\mathrm{BH}}$-M$_{\star}$, M$_{\mathrm{BH}}$-$\sigma_\star$ and the baryonic Tully-Fisher relation. We find that the global star formation (SF) of galaxies run with AGN is reduced compared to the one in which  AGN has been turned off, suggesting that this type of feedback is a viable way of suppressing SF in low-mass dwarfs, even though none of our galaxies is completely quenched by $z$=$0$. Furthermore, we found a tight correlation between the median SF rates and the black-hole-to-stellar mass ratio (M$_{\mathrm{BH}}$/M$_{\star}$) in our simulated dwarfs. SF is suppressed due to gas heating in the vicinity of the AGN: less HI gas is available in AGN runs, though the total amount of gas is preserved across the two settings within each galaxy. This indicates that the main effect of AGN feedback in our dwarfs is to heat up and push the gas away from the galaxy's centre rather than expelling it completely. Finally, we show that the two galaxies harbouring the largest BHs have suffered a  considerable (up to $\sim$65$\,\%$) reduction in their central dark matter density, pinpointing the role of AGNs in determining the final dark matter mass distribution within dwarf galaxies. This pilot paper highlights the importance of modelling AGN feedback at the lowest mass scales and the impact this can have on dwarf galaxy evolution.}
   {}


   \maketitle
%

\section{Introduction}\label{sec:INTRO}

The $\Lambda$ Cold Dark Matter ($\Lambda$CDM) model of cosmology is the prevailing theoretical framework for cosmological structure formation, including the distribution and demographics of galaxies \citep[]{Davis1985}. In this cosmology, dark matter (DM) haloes collapse from initial primordial density fluctuations. These haloes then proceed to merge with one another to build larger haloes with sufficient gravity to provoke the collapse and cooling of gas, which produces stars, thereby populating DM  haloes with luminous galaxies. This is the intersection of cosmology and galaxy formation, where key tests lie in understanding the relation between galaxies and DM haloes.\\

Abundance matching tells us that galaxy formation is most efficient for L* galaxies (with halo masses of around $\sim$$10^{12}$$\,\rm M_{\odot}$) and less efficient towards the high-mass and low-mass ends of the galaxy stellar mass function \citep[e.g.][]{Guo2010,Moster2013}. The most widely accepted explanations of these trends invoke some baryonic process. For massive galaxies, energetic feedback from AGN is critical in regulating star formation (SF) \citep[e.g.]{Binney1995,Ferrarese2000,DiMatteo2005,Springel2005b} by heating the interstellar gas and driving galactic winds. Indeed, several numerical simulations have shown that AGN feedback could quench  massive galaxies and bring their properties in agreement with  observations \citep[e.g.][]{Springel2010,Dubois2014,Vogelsberger2014a,Schaye2015,Weinberger2017,Henden2018}.\\

At the low-mass end, reionization \citep[e.g.][]{Somerville2002,Okamoto2008,Pawlik2009} and supernova (SN) feedback \citep[e.g.][]{Larson1974,Dekel1986,White1991} are thought to play a key role in suppressing SF in such small, dwarf galaxies. However, even considering these feedback processes,  observations of dwarf galaxies, particularly within the Local Group, still present several challenges to the $\Lambda$CDM paradigm. Recent examples include: the missing satellite problem, \citep[e.g.][]{Kauffmann1993,Klypin1999,Moore1999a,Bullock2010}; the `too-big-to-fail' problem \citep[e.g.][]{Read2006,BoylanKolchin2012,GarrisonKimmel2014}; the `cusp-core' problem \citep[e.g.][]{Flores1994,Moore1994,deBlok2010}; and the  diversity of rotation curves problem \citep[e.g.][]{Oman2015,SantosSantos2018,SantosSantos2020}.\\

Some of these problems can be  partially solved by the inclusion of properly modelled baryonic physics in simulations, such as the missing satellites problem, that can be explained by  feedback processes reducing the efficiency of dwarf galaxy formation \cite[e.g.][]{Sawala2016}, or the `cusp-core' problem, that can likewise be alleviated if SNe driven outflows flatten the central DM cusps of simulated dwarfs \citep{Governato2010,Pontzen2012,DiCintio2014, Brook2015}.\\

Other issues are, however, still being actively scrutinized in the literature: despite three decades  of effort, a comprehensive, self-consistent solution to all of these dwarf galaxy problems remains elusive
\citep[see][for a discussion on the historical and new tensions of the $\Lambda$CDM model]{Sales2022}.
An attractive avenue, as suggested by recent theoretical works  \citep{Silk2017,Dashyan2018}, is to consider the role and influence of AGN feedback in low-mass galaxies, which might provide a unifying scheme to explain most of the above-mentioned problems.
 In particular, \citet{Silk2017} postulates that massive BHs could be ubiquitously present in all early-forming dwarfs, having been active in their gas-rich past, but being mostly passive today. Could AGN feedback play an important role in the evolution of dwarf galaxies, as well as of massive ones? Over the past years, mounting observational evidence pointed to the fact that the number of dwarf galaxies hosting AGNs could be larger than previously thought (see \cite{Mezcua2017} for a recent review on the topic).\\

Several observational techniques have been employed to look for AGNs signatures in dwarf galaxies, finding a non-negligible occupation fraction in most cases: these studies include detections based on H$_\alpha$ emission lines \citep{Greene2004}, X-ray observations \citep[e.g.][]{Reines2013,Baldassare2015,Baldassare2017} and optical emission \citep[e.g.][]{Greene2004,Greene2007,Reines2013,Chilingarian2018}. Furthermore,  observations in optical, X-ray, or even infrared (IR) using integral-field unit (IFU) spectroscopic surveys, such as MANGA/SDSS \citep{Bundy2015}, have identified several "hidden" AGN in dwarfs \citep[e.g.] []{Mezcua2016,Penny2018, Baldassare2018,Mezcua2018a,Mezcua2018b,Mezcua2019,Dickey2019,Mezcua2020,Birchall2020,Baldassare2020}. However, a systematic study of AGNs at low-masses using IFU data is still missing.\\

The discovery of AGNs in dwarf galaxies has motivated, in recent years, the study of the impact of central supermassive black holes (SMBHs) on the host dwarf galaxies by using hydrodynamical simulations. Large-scale simulations are useful to explore the AGNs' occupation fractions at dwarf scales. However, the current literature on the subject is characterized by significant variations in findings: while some simulations overproduce bright AGNs, such as \texttt{ROMULUS} and \texttt{IllustrisTNG} \citep{Sharma2020,Haidar2022}, others such as \texttt{EAGLE}, \texttt{Illustris}, or \texttt{FABLE} underproduce bright AGNs \citep[e.g.][]{Koudmani2021,Haidar2022}.  Indeed, the use of different feedback schemes can produce significant discrepancies between simulations \citep{Habouzit2017}. Several studies have shown how strong SN feedback could even hinder BH growth by preventing the accretion of gas \citep[e.g.][]{Dubois2015,Habouzit2017,Angles2017,Trebitsch2018,Koudmani2022}. In those cases in which  BHs are instead able to grow and accrete gas, the related AGNs can generate outflows that can  play a determinant role in  regulating  SF in dwarf galaxies \citep[e.g.][]{Koudmani2019,Barai2019,Sharma2020,Koudmani2021,Koudmani2022}.\\

Nevertheless, the degree to which AGN feedback is able to suppress and quench SF in dwarfs effectively is still an open question: the field has now reached a stage at which a comprehensive investigation of this issue is not only warranted but also needed.\\

In this paper, we use zoom-in simulations of dwarf galaxies to study the impact of AGN feedback on galaxy evolution at the low-mass end, by running exactly the same cosmological initial conditions of dwarfs with and without the inclusion of AGNs. In section \ref{sec:auriga}, we briefly describe the modified \texttt{AURIGA} simulations used in this work. The BH seeding and AGN feedback modelling are described in \ref{sec:BHseed}. The properties of the galaxy sample are explained in section \ref{sec:sample}. In section \ref{sec:Result} we present scaling relations relative to our sample hosting AGNs (section \ref{sec:scaling}) as well as the impact of AGNs on global properties of simulated galaxies, such as the stellar-to-halo mass relation (section \ref{sec:SHMR}), the Baryonic Tully-Fisher relation (\ref{sec:BTFR}) and the dependence on the SF history (SFH) (section \ref{sec:SFH}). The DM profiles' dependence on AGNs is described in section \ref{sec:DMprof}. We touch upon gas properties and how they are affected by the inclusion of AGNs in section \ref{sec:HIproperties}, a thorough investigation of which we defer to a future dedicated study. We discuss our results in section \ref{sec:Conclusions}.\\

\section{Cosmological magneto-hydrodynamic dwarf galaxy simulations}\label{sec:Methodology}
\subsection{Simulation code and physics model}\label{sec:auriga}

\begin{table*}
    \centering
        \caption{$z$=$0$ properties of simulated dwarf galaxies run with and without the inclusion of AGNs. Columns represent: (1) AURIGA ID; (2) run name; for the sample with AGN: (3) halo virial mass, (4) halo virial radius, (5) stellar mass, (6) BH mass; and for the sample without AGN: (7) halo virial mass, (8) halo virial radius, (9) stellar mass.}
    \begin{tabular}{llccccccc}
    & \multicolumn{4}{c}{AGN} & \multicolumn{3}{c}{no AGN}\\
    \cmidrule(lr){3-6}
    \cmidrule(lr){7-9}
      {ID}& {Run}  & $\frac{\rm M_{200}}{(10^{10} \rm M_\odot)}$ & $\frac{ \rm R_{200}}{(\mathrm{kpc})}$ & $\frac{\rm M_\star}{(10^{9}\rm M_\odot)}$& $\frac{\rm M_{\mathrm{BH}}}{(10^{6} \rm M_\odot)}$ & $\frac{\rm M_{200}}{(10^{10}\rm M_\odot)}$ & $\frac{\rm R_{200}}{(\mathrm{kpc})}$ & $\frac{\rm M_\star}{(10^{9}\rm M_\odot)}$\\
       \midrule
   
       Au5 & H0  &  27.44  &  136.97  &  5.70  &  13.34  &  28.99  &  139.51   &  6.58  \\
       Au3 & H1  &  21.50  &  126.27  &  4.21  &  9.37  &  22.53  &  128.25  &  5.89  \\
       Au1 & H2  &  20.31  &  123.91  &  2.74  &  4.18  &  21.08  &  125.45  &  3.70  \\
       Au4 & H3  &  28.62  &  138.90  &  2.89  &  2.76  &  29.30  &  140.00  &  3.65  \\
       Au2 & H4  &  12.41  &  105.13  &  0.96  &  1.97  &  12.94  &  106.60  &  1.88  \\
       Au0 & H5  &  9.37  &  95.73  &  0.69  &  1.80  &  9.63  &  96.61  &  1.05  \\
       Au7 & H6  &  13.84  &  109.03  &  1.13  &  1.67  &  14.77  &  111.41  &  1.82  \\
       Au11 & H7 &  7.52  &  88.96  &  0.64  &  1.14  &  7.90  &  90.43  &  0.90  \\
       Au6 & H8  &  7.78  &  90.00  &  0.78  &  0.88  &  7.93  &  90.56  &  0.86  \\
       Au10 & H9  &  6.54  &  84.93  &  0.41  &  0.41  &  6.71  &  85.66  &  0.50  \\
       Au8 & H10  &  9.67  &  96.74  &  0.36  &  0.41  &  9.84  &  97.31  &  0.41  \\
       Au9 & H11  &  8.85  &  93.92  &  0.24  &  0.29  &  8.92  &  94.19  &  0.25  \\
       \midrule

    \end{tabular}\\
    \raggedright
    \label{tab1}
\end{table*}

The simulations in this study were performed with the massively parallel N-body, second-order accurate magnetohydrodynamics (MHD) code \texttt{AREPO} \citep[]{Springel2010,Pakmor15}. \texttt{AREPO} calculates gravitational forces using a TreePM method, which incorporates a fast Fourier Transform for long-range forces and a hierarchical oct-tree algorithm for short-range forces. The code utilizes a dynamic, unstructured mesh constructed through Voronoi tessellation, allowing for the finite-volume discretisation of the MHD equations.\\

The simulations include a version of the \texttt{AURIGA} physics model, which includes: primordial and metal line cooling; a uniform UV background that completes reionisation at $z$=$6$; the \citep{SH03} subgrid model for SF; magnetic fields seeded at initial conditions with a uniform distribution in a random orientation \citep{PGG17}; and energetic feedback from AGN and supernovae type II (SNII). Each star particle represents a single stellar population characterized by a mass, metallicity, and age. Mass loss and chemical enrichment are modelled from stellar evolutionary processes, specifically type Ia supernovae (SNIa), Asymptotic Giant Branch (AGB) stars and type II supernovae (SNII). The \texttt{AURIGA} physics model is fully described in \citet{Grand2017}. \\

To model galactic winds driven by SNII, wind particles are launched with a velocity that scales with the local, one-dimensional DM  velocity dispersion, $\sigma_{\rm DM}$, with an imposed minimum wind velocity of $v_{\rm w,min}$=350$\,\rm km\,s^{-1}$ following \citep{Annalisa2018}, instead of the default value of $v_{\rm w,min}$=0$\,\rm km\,s^{-1}$ used in the original \texttt{AURIGA} physics model. The non-zero minimum wind velocity that we use in this paper has the effect of reducing the wind mass loading factor and the total stellar mass for haloes with masses $\leq$10$^{11}\,\rm M_{\odot}$ compared to the original \texttt{AURIGA} physics model. As discussed in section \ref{sec:SHMR}, this shifts the galaxies onto or just below the stellar-mass-halo-mass relation, compared to the original case of no minimum wind velocity for which the galaxies lie slightly above it.\\

\subsubsection{Black Hole seeding, accretion \& feedback}\label{sec:BHseed}

BHs are seeded with a mass of $10^5$${\,\rm M_{\odot}}\,h^{-1}$ in halo FOF groups of masses greater than M$_{\rm FOF}$=5$\times$$10^{10}$${\,\rm M_\odot}\,h^{-1}$ and are placed at the position of the densest gas cell. BH particles act as sinks that draw in mass from the nearest neighbour gas cells \citep{Springel2005}. The rate of accretion follows the Eddington-limited Bondi–Hoyle–Lyttleton accretion formula \citep[][]{Bondi1944,Bondi1952} in addition to a term that models the radio mode accretion, given by

\begin{equation}
    \Dot{M}_{\mathrm{BH}} = \mathrm{min}\left[\frac{4\pi G^2 M^2_{\mathrm{BH}} \rho}{\left( c_s^2 + v^2_{\mathrm{BH}}\right)^{3/2}} + \frac{R(T,z)L_{\mathrm{X}}}{\epsilon_\mathrm{f}\epsilon_\mathrm{r}c^2}, \Dot{M}_{\mathrm{Edd}}\right]
\end{equation}

where $\rho$ and $c_s$ are the density and sound speed of the surrounding gas, $v_{\mathrm{BH}}$ is the velocity of the BH relative to the gas, $\Dot{M}_{\mathrm{Edd}}$ is the Eddington accretion rate, $L_\mathrm{X}$ is calculated from the thermal state and cooling time of the non-star-forming gas cells and $R(T,z)$ is a scaling factor calculated from relations presented in \cite{Nulsen2000}. We define the BH radiative efficiency parameter to be $\epsilon_\mathrm{r}$=$0.2$ and the fraction of released energy that couples thermally to the gas to be $\epsilon_\mathrm{f}$=$0.07$.

The radio mode term comes from the assumption that the hot halo gas is in thermodynamic equilibrium so that X-ray losses are compensated by thermal energy injection via the gentle inflation of multiple bubbles within the virial radius of the halo \citep[see][]{Grand2017}. However, this mode of AGN feedback is important for more massive haloes like Milky Way-mass spiral galaxies and giant ellipticals; radio mode feedback does not activate in any of our simulations. Therefore, BH  feedback is, in practice, that of the quasar mode, which is modelled via isotropic thermal energy injection into neighbouring gas cells, where the total energy to be injected is given by:

\begin{equation}
    \label{eq:AGNener}
    \Dot{E} = \epsilon_\mathrm{f}\epsilon_\mathrm{r}\Dot{M}_{\mathrm{BH}}c^2,
\end{equation}
where $c$ is the speed of light. The thermal energy is injected into neighbouring gas cells with the amount of energy per cell following an inverse square relation. The additional radiative feedback adds to the UV background locally (see \cite{Vogelsberger2013} for more details).

\subsection{Initial conditions}\label{sec:sample}
Our simulated haloes are selected from the DM-only counterpart of the \texttt{EAGLE} simulation, in a comoving box with a side-length of 67.77$\,h^{-1}\,\rm Mpc$ (L100N1504) as introduced in \citet{Schaye2015}. We select 12 isolated haloes with a $z$=$0$ mass between $5\times10^{10}\,\rm M_{\odot}$ and 5$\times$10$^{11}\,\rm M_{\odot}$ using the same 
isolation criterion described in \citet{Grand2017}\footnote{Zoom-in simulations of these objects run with the original \texttt{AURIGA} physics model are publicly available as described in \citet{Grand2024}.}. We then re-simulate these haloes using the zoom-in technique, adopting the cosmological parameters given in \cite{Planck2014}: $\Omega _m$=$0.307$, $\Omega _b$=$0.048$, $\Omega _{\Lambda}$=$0.693$, $\sigma_8$=$0.8288$, and a Hubble constant of $H_0$=$100\,h\,\rm km\,s^{-1}\,\rm Mpc^{-1}$, where $h$=$0.6777$. The typical mass resolution contained in each Lagrangian volume for gas and DM particles is $m_{\rm gas}$=5$\times$$10^4$$\,\rm M_\odot$ and $m_{ \rm DM}$=3$\times$$10^5$$\,\rm M_\odot$, respectively. The comoving softening length of collisionless particles is set to 500$\,h^{-1}\,\rm cpc$; the physical softening length is kept fixed to 250$\,h^{-1}\,\rm pc$ below $z$=$1$.

Haloes and subhalos are identified using the Amiga Halo Finder, \texttt{AHF} \citep[][]{Knollmann2009}, in which halo virial masses,  M$_{200}$, are defined as the masses contained within a sphere of virial radius R$_{200}$, enclosing  $\Delta_{200}$$\simeq$200 times the critical density of the Universe at $z$=$0$. Subhalos are considered resolved if they have at least 200 particles. The central halo is the one found at the minimum gravitational potential of the group. All other subhalos in the same group are categorized as satellites of the central one. For this work, only the central halo has been considered to focus on internal processes of the galaxy, rather than on environmental ones. The bulk of the analysis is done using a modified version of \texttt{PYNBODY} \citep[][]{Pontzen2013} that is compatible with \texttt{AURIGA}.


We perform two sets of simulations: the first one uses the physics model described in section~\ref{sec:auriga}. We hereafter refer to this configuration as \textit{fiducial} or \textit{AGN} run. In the second configuration, we run exactly the same galaxies but without the inclusion of BHs and related AGN feedback. We will refer to this configuration as the \textit{no-AGN} run. Table~\ref{tab1} summarises the final halo, stellar and BH masses, as well as the halo virial radius, for our simulated galaxies run with and without the inclusion of AGNs.

\begin{figure}
	\includegraphics[width=\columnwidth]{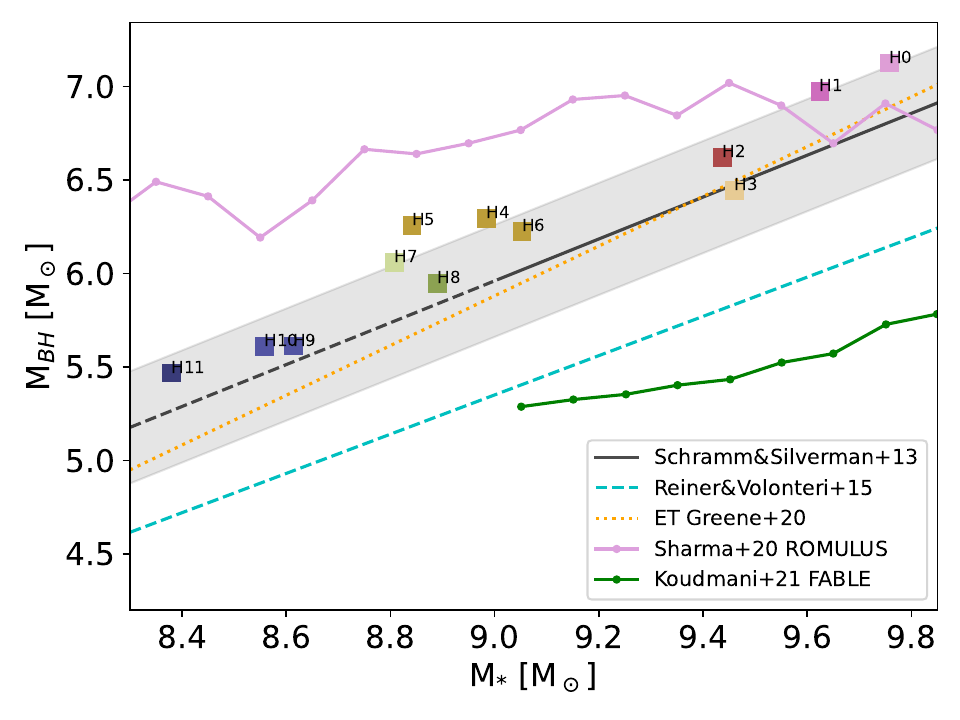}
    \caption{BH mass-stellar mass relation for the 12  \texttt{AURIGA} dwarf galaxies run with the fiducial configuration. Each galaxy is shown in a different colour to facilitate comparisons with the figures in the following sections. The black solid line shows the median relation given by \protect\cite{Schramm2013}, with the dashed line indicating the range below which such relation is extrapolated and the shaded area showing a $0.2$ dex scatter. The dashed cyan line shows the relation of \protect\cite{Reines2015}, while the orange dotted line indicates the same relation for \protect\cite{Greene2020} Early-Type galaxies. For comparison with ongoing simulation results, we show in green 
     the mean M$_{\rm BH}$-M$_{\star}$ relation from the \texttt{FABLE} simulations suite \citep{Koudmani2021} and in purple that  from the \texttt{ROMULUS} simulation \citep{Sharma2020}.}

    \label{fig:Mstar-Mbh}
\end{figure}

\section{Results}\label{sec:Result}

Our work aims to understand the impact of AGN feedback on the properties of intermediate-mass dwarf galaxies by analysing and comparing our fiducial simulation set with the no-AGN runs. In this section, we present scaling relations directly connected to the presence of a central SMBH in the fiducial run, before studying how AGNs affect various galaxy properties.

\subsection{The M$_{\rm BH}$-M$_{\star}$ and M$_{\rm BH}$-$\sigma_{\star}$ relations}\label{sec:scaling}

\begin{figure}
	\includegraphics[width=\columnwidth]{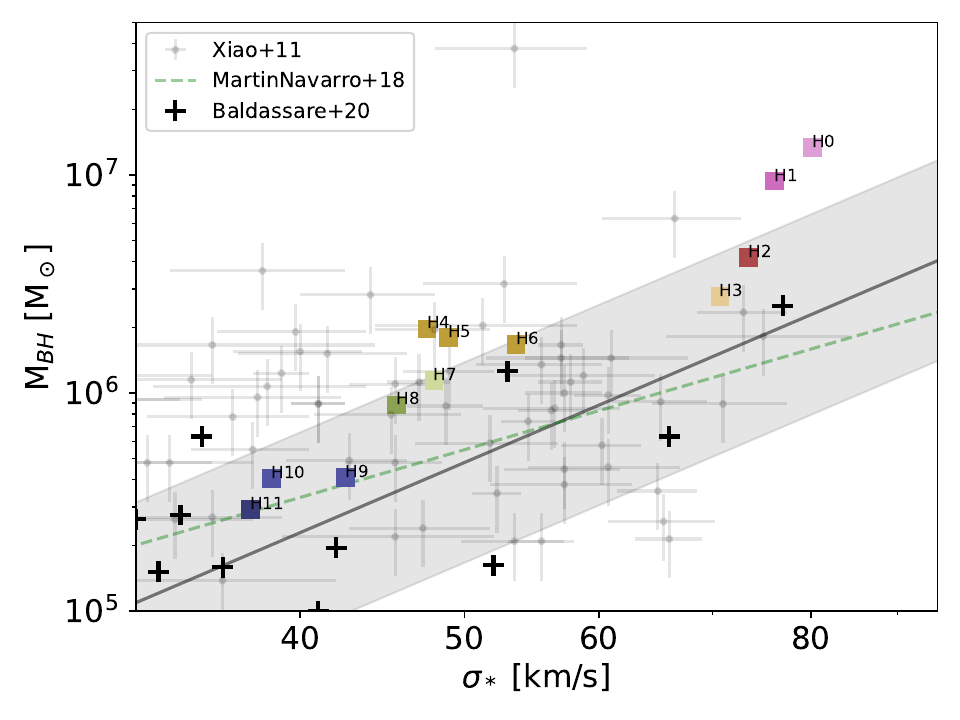}
    \caption{BH mass versus stellar velocity dispersion ($\sigma_{\star}$) relation, for our simulated dwarfs indicated as coloured squares. The velocity dispersion of the stellar component has been measured inside 0.5$\,\rm R_{\mathrm{eff}}$ for each galaxy. As grey points with error bars, observations from \protect\cite{Xiao2011}  with a 0.2 dex scatter as grey shaded area. As black crosses, observations from \citet{Baldassare2020}. The green line shows the \protect\cite{MartinNavarro2018}  observational relation obtained for low-mass Seyfert 1 galaxies. All of our simulated galaxies follow fairly well current existing observational relations between the mass of the central BH and the velocity dispersion of the galaxy's stellar component.}
    \label{fig:sigma-Mbh}
\end{figure}

In Fig. \ref{fig:Mstar-Mbh}, we examine the correlation between the mass of the central BH  and the stellar mass of the host galaxy, at $z$=$0$. We found that our simulations follow the positive trends obtained in observations as reported in \citet{Schramm2013} (dark line) as well as in \citet{Greene2020} for early-type galaxies (orange dotted line): galaxies with progressively smaller stellar masses host less and less massive BHs at their centre, all the way down to the smallest scales simulated here. Our simulations fall nicely within the scatter of the \cite{Schramm2013} relation. We further show the same scaling relations as obtained by \cite{Reines2015}, in cyan, who used a sample of broad-line AGNs in the nearby universe, constructed mainly using SDSS spectroscopy, looking for Seyfert-like narrow-line ratios and broad H$_\alpha$ emission.

To compare with recent theoretical works, we additionally show the mean M$_{\rm BH}$-M$_{\star}$ relation from the \texttt{FABLE} simulations  \citep[green line]{Koudmani2021}, which also use the \texttt{AREPO} code but with a different galaxy formation model from \texttt{AURIGA}, and from the \texttt{ROMULUS25} simulations \citep[purple line]{Sharma2020}, that employ the code ChaNGa \citep{Menon2015} with baryonic prescriptions from Gasoline2 \citep{Stinson2006,Shen2010,Wadsley2017}. Interestingly, across the galaxy stellar mass range explored, the  \texttt{FABLE} simulations produce a population of BHs that are under-massive compared to both our simulation results and to currently observed relations. In contrast, the \texttt{ROMULUS25} suite produces over-massive BHs that deviate from the M$_{\rm BH}$-M$_{\star}$ relation at small scales. This highlights the strong dependence of the final BH mass-stellar mass relation on the BH seeding mass and accretion model employed in each case,  the former being 10$^5$$\, \rm M_\odot$ and 10$^6$$\,\rm M_{\odot}$ for the \texttt{FABLE} and \texttt{ROMULUS25} simulations, respectively.

\begin{figure}
	\includegraphics[width=\columnwidth]{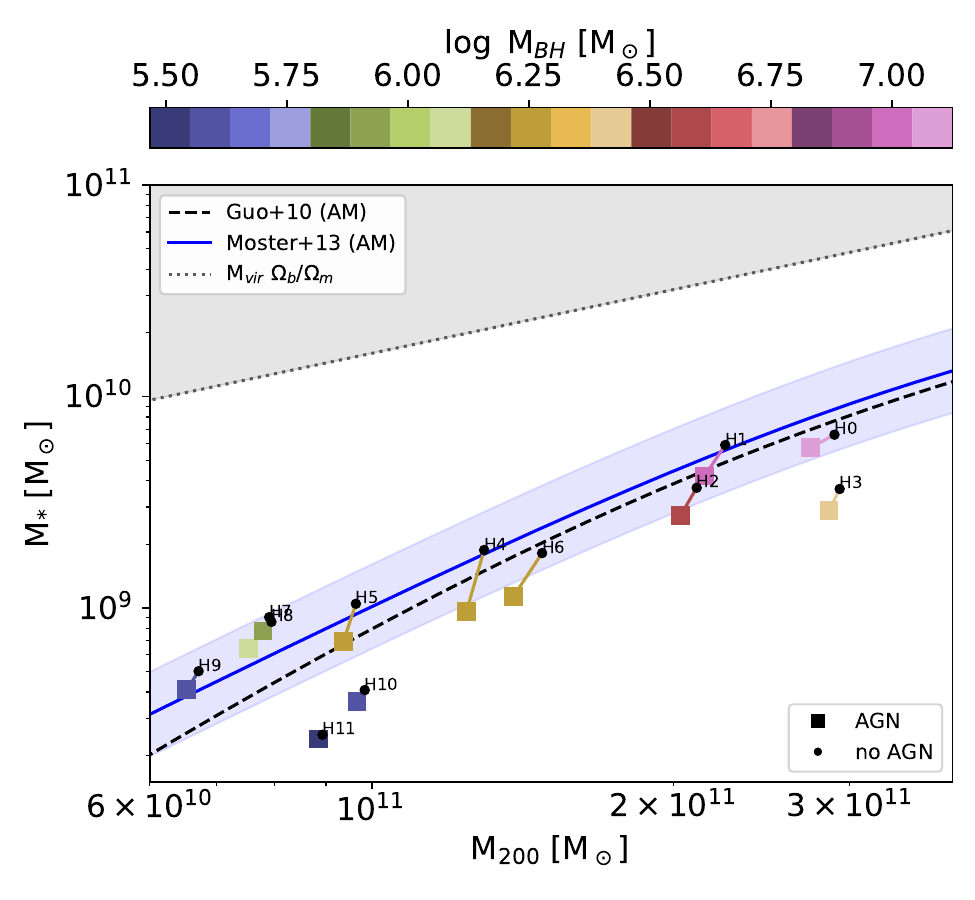}
    \caption{Stellar-to-halo mass relation of simulated dwarf galaxies: black points represent galaxies run without BHs, while coloured squares those galaxies run in the fiducial model, i.e. including a central BH. In this case, each galaxy is shown with a different colour reflecting the mass of its central BH. 
  Each fiducial/no-AGN pair (i.e. a simulation with the exact same initial conditions) is connected with a line. In solid blue, the abundance matching relation of \protect\cite{Moster2013} with a scatter of 0.2 dex as shaded blue region. As dashed dark line,   estimates by \protect\cite{Guo2010}. Galaxies harbouring BHs more massive than $\sim$10$^6$$\,\rm M_{\odot}$ have a  strongly reduced M$_{\star}$ compared to their no-AGN counterparts.}
    \label{fig:SHMR}
\end{figure}

In Fig. \ref{fig:sigma-Mbh}, we show the relation between BH mass and central velocity dispersion of the host galaxy's stellar component,
and compare our simulation results with observational works that
 explored the low-mass end of such M$_{\rm BH}$-$\sigma_{\star}$ relation.
While this relation has been studied in detail in the high-mass regime, only in recent years has this been extended to the low-mass range. As in observations, we define $\sigma_{\star}$ as the average stellar velocity dispersion within half of the effective radius, R$_{\rm eff}$, in each galaxy.   Our sample follows reasonably well the trend derived in  \cite{Xiao2011} (grey points) and later on in \citet{Baldassare2020} (black crosses) within the observed scatter (grey shaded area). Further, we compare with the relation obtained in \citet{MartinNavarro2018} for low-mass Seyfert 1 galaxies, in which they compute the M$_{\rm BH}$ following the same assumption of \cite{Baldassare2020} and \cite{Xiao2011} that the gas is virialized, using the luminosity and the FWHM of the broad H$_\alpha$ component (see eq. 1 from \citealt{MartinNavarro2018} and eq. 6 from \citealt{Xiao2011}). The agreement between our sample and the observed M$_{\rm BH}$-M$_{\star}$ and M$_{\rm BH}$-$\sigma_{\star}$ scaling relations make our simulation suite the ideal starting point to explore the impact of AGN feedback on dwarf galaxies.

In the next sections we study the stellar-to-halo mass relation, the baryonic Tully-Fisher relation and the SFHs of our simulated dwarfs with and without the inclusion of AGN feedback, and compare these two settings with observational data.

\subsection{The stellar-to-halo mass relation}\label{sec:SHMR}

The relation between the stellar and the virial mass of simulated dwarfs in the fiducial and no-AGN configurations can be seen in Fig. \ref{fig:SHMR}.  Squares represent individual galaxies in the fiducial run, colour-coded according to their central BH mass. Fiducial simulations are linked to their no-AGN galaxy counterparts, denoted by black dots. 
This image shows that galaxies in both configurations agree fairly well with the semi-empirical M$_{\star}$-M$_{200}$ relation of \citet{Moster2013} and \citet{Guo2010} within the allowed scatter. However, it can be noticed that those galaxies harbouring a central BH of mass larger than $\sim$$10^6$$\,\rm M_{\odot}$ have a strongly reduced total stellar mass compared to their no-AGN counterpart. This suggests that AGNs have a non-negligible impact on the global SFH of dwarf galaxies, being able to reduce the stellar mass of our simulated galaxies by as much as a factor of 2 (see for example galaxy H4 in table \ref{tab1}). On the contrary, the impact of AGN feedback is negligible (though still present) for those galaxies hosting a BH less massive than $\sim$$10^6$$\,\rm M_{\odot}$, which are typically found in galaxies with stellar masses  M$_{\star}$$\leq$$10^{8.8}$$\,\rm M_{\odot}$.

\begin{figure}
	\includegraphics[width=\columnwidth]{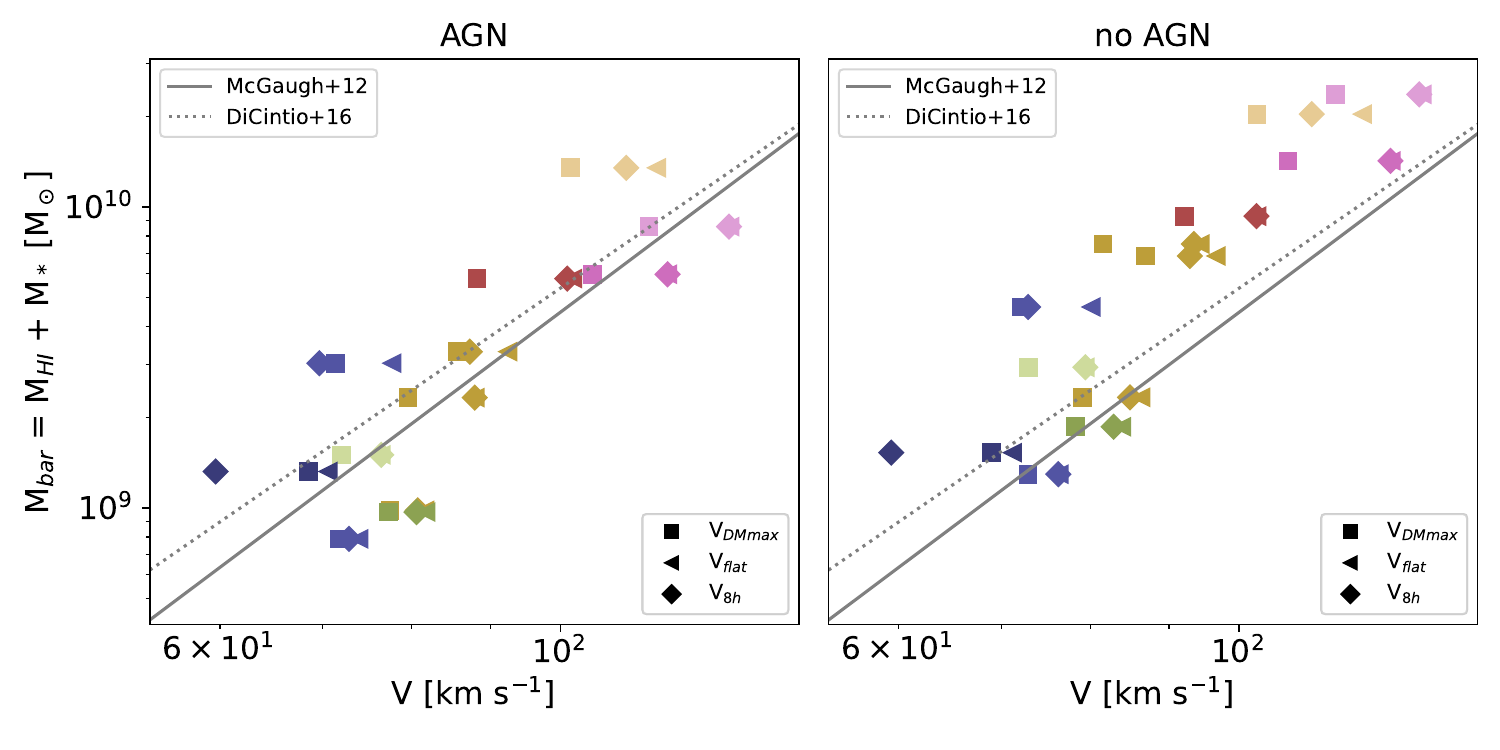}
    \caption{The Baryonic Tully-Fisher relation of simulated dwarf galaxies, run with (left panel) and without (right panel) the inclusion of BH and AGN feedback. Each galaxy is shown in a different colour according to its $z$=$0$ BH mass, as in Fig. \ref{fig:SHMR}.  Different symbols indicate different ways of measuring the rotational velocity, $V_{\mathrm{rot}}$: V$_{\mathrm{DM,max}}$, V$_{\mathrm{flat}}$ and V$_{\rm 8h}$ are shown as squares, triangles and diamonds, respectively.  As the solid line, the observational relations given by \protect\cite{McGaugh2012}, and as the dotted line, the relation obtained in  \protect\cite{DiCintio2016} using semi-analytic models. Simulated galaxies that include AGN feedback are in better agreement with the observed BTFR.}
    \label{fig:BTFR}
\end{figure}

\subsection{The Baryonic Tully-Fisher relation}\label{sec:BTFR}

To infer whether our simulated galaxies represent a trustworthy sample, we study in detail their baryonic Tully-Fisher relation (BTFR), i.e. the relationship between the rotation velocity of galaxies, as measured by their HI line, and their baryonic mass content \citep{McGaugh2012}. As neutral atomic gas dominates the gas component in disk galaxies, in this analysis we define the baryonic mass of a galaxy as the sum of the neutral hydrogen plus the stellar mass, i.e. M$_{\rm b}$=M$_{\star}$+M$_{\mathrm{HI}}$, in line with observations.
While the slope of the BTFR is well constrained at high masses, it suffers from several observational uncertainties at lower, dwarf-mass scales. Historically, the velocity of the galaxies was measured at the radius at which the rotation curve reaches its flat part, $V_{\rm flat}$ \citep[]{Stark2009}. However, for dwarf galaxies especially, this point is often not reached at the outermost measured radius. 
\citet{Brook16}, using simulations, highlighted how the slope of the BTFR changes at low masses when measuring the rotation velocity of the galaxy at different positions, such as at $V_{\rm flat}$, or at the extent of the HI gas disk, $V_{\rm last}$, or using the width of
the HI line profiles.

Keeping this in mind, here we will use three different definitions to measure the rotation velocity of our simulated dwarfs, and compare them with BTFRs presented in the literature: 

\begin{itemize}
    \item V$_{\mathrm{DM},\mathrm{max}}$: the maximum circular velocity of the DM halo of each galaxy;
    \item V$_{\mathrm{flat}}$: the circular velocity at the flat part of the rotation curve, as done in \citep{McGaugh2012} for observed late-type galaxies; 
    \item V$_{\rm 8h}$: the rotation velocity at eight times the disk radii, R$_{\mathrm{disc}}$, such that R$_{\mathrm{eff}}$=1.678$\times$R$_{\mathrm{disc}}$, radius at which most of the baryonic mass is found, as done in \citep[]{DiCintio2016} using semi-analytic models.
\end{itemize}

\begin{figure*}
  \centering
  \includegraphics[width=\textwidth]{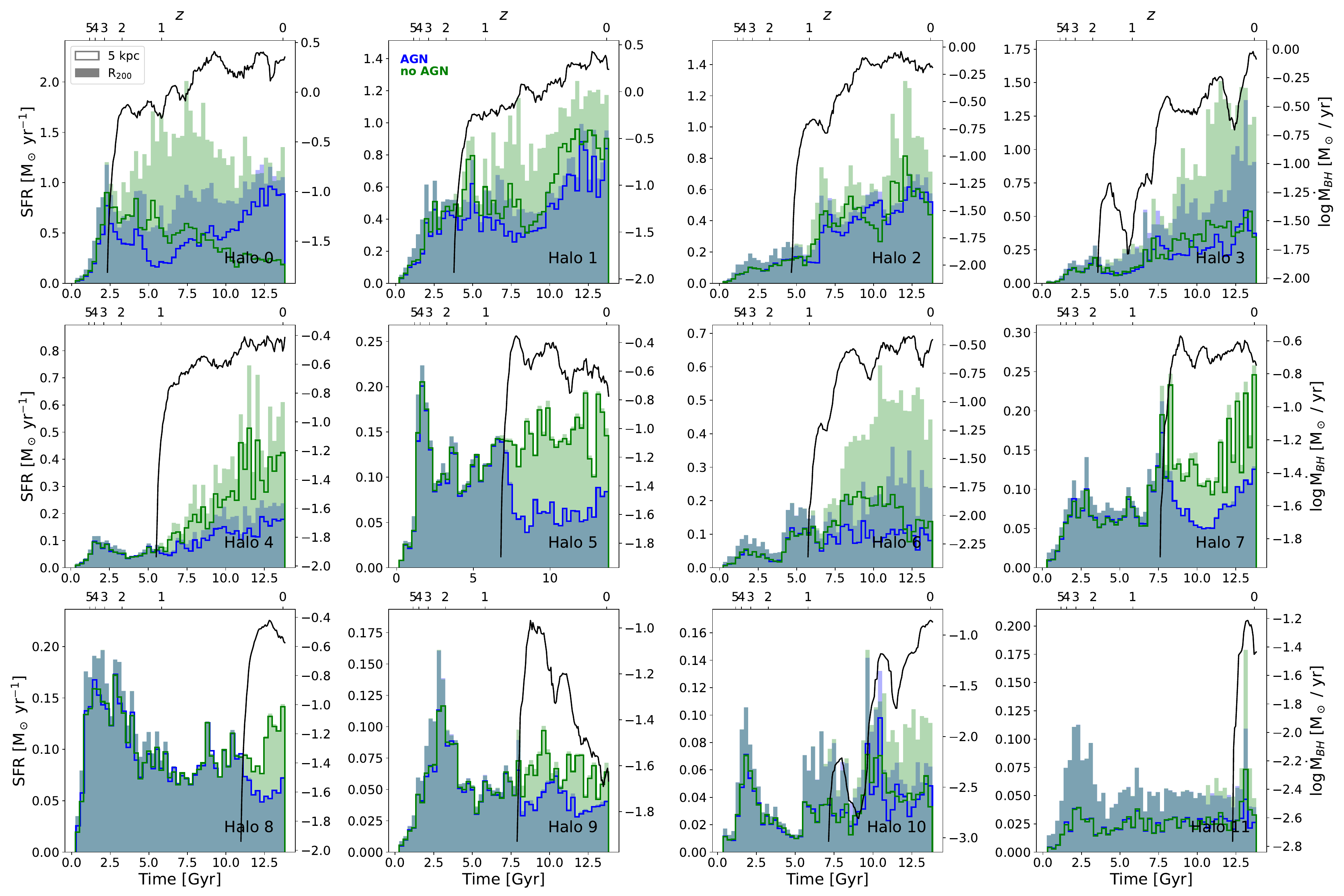}
  \caption{SFHs of simulated dwarf galaxies, within the virial radius (shaded histograms) and within 5 kpc from the centre (open histograms).  Galaxies that evolved with AGN feedback are represented in blue, while those without AGN are shown in green, and they are ordered according to their $z$=$0$ BH mass (most massive BH hosted in Halo 0, top left, least massive in Halo 11, bottom right). The logarithm of the BH accretion rate through time is indicated on the right-hand Y-axis in each panel and is shown as a black line. }
  \label{fig:SFH}
\end{figure*}

 We measure the amount of neutral atomic hydrogen in the simulations following the phenomenological method described in \citet{Marinacci2017} and based on \citet{leroy08}. This method consists of fitting the ratio between the column density of molecular over atomic hydrogen with a functional form depending on the gas mid-plane pressure, to then compute the atomic HI fraction. We follow such an empirical approach since the galaxy formation modules currently employed in \texttt{AURIGA} do not account for the mechanisms responsible for the creation and destruction of molecular hydrogen.
 
 In Fig.~\ref{fig:BTFR}, we show the BTFR derived for our set of simulated galaxies using the three definitions of velocity as above: each symbol indicates a particular way of measuring $V_{\mathrm{rot}}$, while each colour represent a specific galaxy, following the same colour scheme as in Fig. \ref{fig:SHMR}, based on the mass of the central BH. The left panel refers to the fiducial, AGN model, while the right panel to the same simulations run without BHs.
Comparing both configurations with observed \citep{McGaugh2012}
 and semi-analytic \citep{DiCintio2016}
 BTFRs, we note that the inclusion of AGN feedback brings most of the galaxies in line with expectations, especially the most massive of our sample. This is because the total HI gas + stellar mass decreases in the AGN configuration. This suggests that, although the presence of a central SMBH does not seem to be essential to reproduce the  M$_{\star}$-M$_{200}$ relation, it is important in order to correctly match the total baryonic mass of galaxies vs their rotational velocity. Dwarf galaxies run without the inclusion of AGN feedback are consistently above such relation, indicating that they retain too much baryonic material compared to observational data.

\subsection{Star formation histories}\label{sec:SFH}

We now proceed to examine how AGN feedback affects the SFH of our set of 12 simulated dwarf galaxies. Fig. \ref{fig:SFH} shows the SF as a function of time,  taking into account all the star particles within the virial radius of each halo (shaded histograms) or by including only those stars within 5 kpc from the galaxy centre (solid line histograms). In both cases, the stars are identified at $z$=$0$ and the corresponding ages are used to derive a SFH.
 Galaxies are ordered by the mass of their central BH: the top-left panel shows the SFH of the dwarfs that host the most massive BH, whereas the bottom-right panel shows the smallest one. In each panel, the logarithm of the BH accretion rate through time is plotted as a solid black line (with rates given on the right-hand axis) as a direct indication of the AGN feedback at different epochs. SFHs of galaxies run with AGNs are shown in blue, while those without AGNs are in green.
 
\begin{figure*}
	\includegraphics[width=\textwidth]{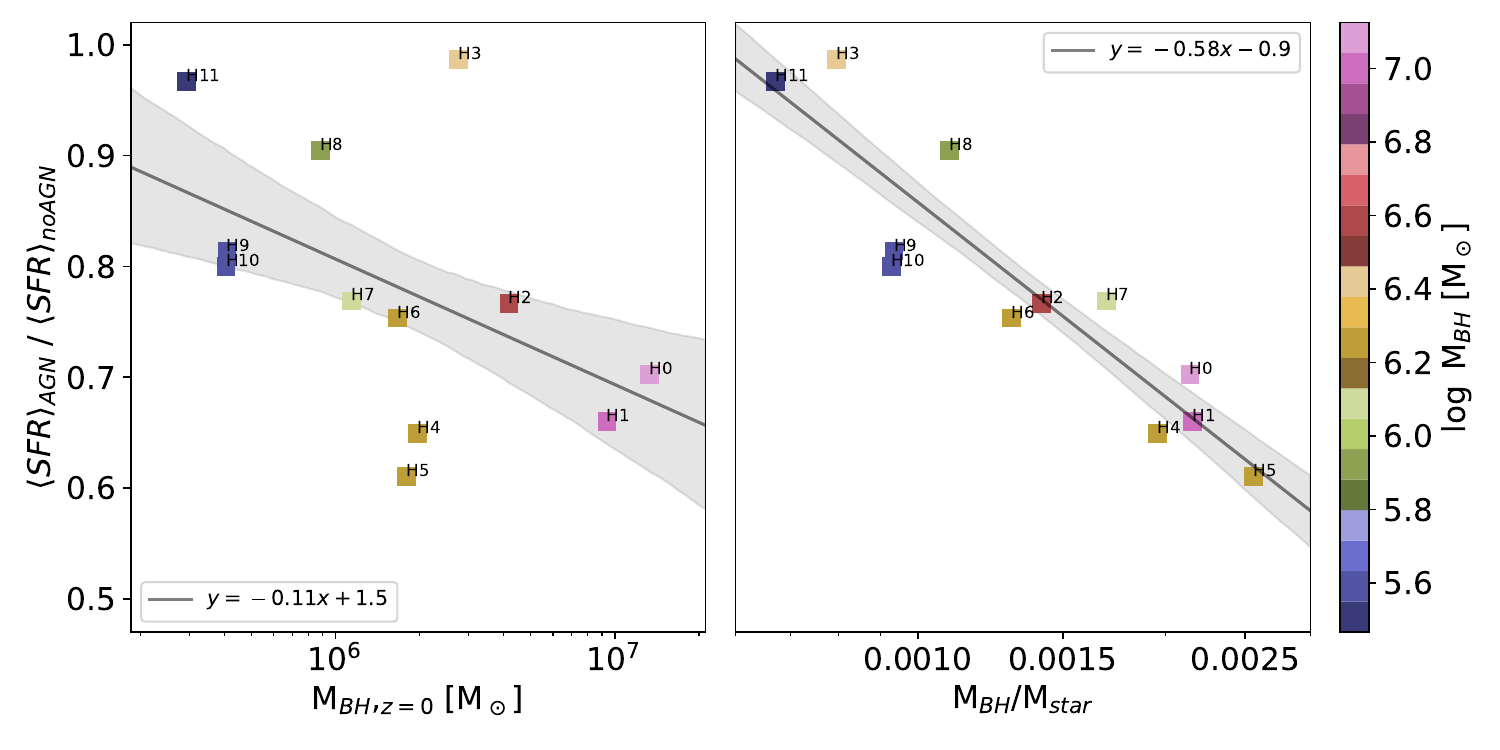}
    \caption{Ratio between the median SFR in the fiducial and non-AGN runs, across cosmic time, for our set of simulated dwarf galaxies, coloured-coded by BH mass. In the left-hand panel, the $\langle \rm SFR \rangle_{\mathrm{AGN}}$/$\langle \rm SFR \rangle_{\mathrm{noAGN}}$  ratio is shown against the $z$=$0$ BH mass of each galaxy in the fiducial run, as a proxy for the total amount of energy available as AGN feedback. In the right-hand panel, this ratio is instead plotted against the BH mass over the stellar mass of the simulated galaxies: this relation shows much less scatter and a stronger correlation. The solid line and grey-shaded region show the best linear fit and the 1$\sigma$ error of the sample, respectively.}
    \label{fig:SFR}
\end{figure*}

Fig. \ref{fig:SFH} shows a consistent trend in which the SF of the dwarfs including AGNs decreases relatively to their no-AGN counterpart, once the central BH is seeded and starts accreting mass. The total amount of suppression in SF appears to scale with the BH accretion rate and with the duration of the epoch in which significant accretion occurs. Thus, the suppression is more evident for the most massive galaxies (at $z$=$0$) that seed their BHs relatively early, such as H0 (top-left panel of Fig.~\ref{fig:SFH}). In contrast, the SFH of the simulated dwarf with the smallest BH (lower-right panel of Fig.~\ref{fig:SFH}) is almost the same in each configuration until the last $\sim$1 Gyr of evolution, when the BH is finally seeded. 
Note that in each halo, the SFHs are almost identical until the first appearance of the BH, clearly indicating that AGN feedback is the effect responsible for the observed differences between the two configurations.

For most of the simulated dwarfs, a reduction in the global SF (star formation within the virial radius of the halo) in the AGN runs, compared to no-AGN runs, is accompanied by a reduction in central SF (within a spherical radius of 5 kpc). However, in some cases, the suppression of SF is only clear when looking beyond the central 5 kpcs: this is the case of the 4 most massive dwarfs, whose central SF is almost unaltered in the AGN vs no-AGN runs. An extreme case is the SFH of dwarf H0, within 5 kpc from the centre and in the last $\sim$3 Gyrs of its evolution, the SFH is higher in the AGN simulation than in its no-AGN companion simulation, which indicates a sort of "positive feedback". We understand this to be driven by an excess of cold gas available at late times, which was prevented from being transformed into stars because early ($t$$\lesssim$7 Gyr) AGN feedback heated it up above the SF temperature threshold. Owing to high-density conditions in such massive dwarf galaxies, cooling times are short enough that such gas can cool back and contribute to late SF in the inner parts of the galaxy. Or, in other words, AGN feedback is not sufficient to suppress the very central SF of massive dwarf galaxies, with masses $\rm M_{200}$>2$\times10^{11}$$\,\rm M_{\sun}$, all the way to $z$=$0$ (even though some early suppression in the inner regions of such galaxies at earlier times is observed), while it is enough to globally suppress their SF. This issue will be explored in a companion paper on HI gas properties in dwarfs harbouring AGNs (see also discussion in Sec.\ref{sec:HIproperties}).\\

In order to quantify the dependence of  SF on AGN feedback, we now take the ratio of the median SF rate (SFR) in the fiducial configuration, $\langle \rm SFR \rangle_{\mathrm{AGN}}$, and the median SFR in the no-AGN simulations, $\langle \rm SFR \rangle_{\mathrm{noAGN}}$, both across cosmic times. Galaxies for which this ratio is close to 1 are not significantly affected by the presence of a central BH, while those with a ratio less than 1 have experienced AGN-induced SF suppression. The left panel of Fig. \ref{fig:SFR} shows the relation between this $\langle \rm SFR\rangle$ ratio and the $z$=$0$ BH mass of the fiducial sample, for each dwarf. One might expect that the reduction of SF due to AGN feedback could directly relate to the mass of the central BH. While this is the case, we note that this trend presents a notable scatter.
In the right-hand panel of Fig. \ref{fig:SFR}, we therefore plot the $\langle \rm SFR \rangle_{\mathrm{AGN}}$/$\langle \rm SFR \rangle_{\mathrm{noAGN}}$ ratio, versus the BH-stellar mass ratio, M$_{\rm BH}$/M$_{\star}$, for the fiducial sample. This relation shows less scatter compared to the one in which only the BH mass is considered, reflecting the importance of taking into account not only the BH mass and related AGN feedback, but also the mass of the galaxy that hosts the BH: dwarf galaxies with a  more pronounced suppression in SF  are not necessarily the ones with the largest BH (see for example the similar amount of SF reduction in dwarfs H0 and H5).  In other words, larger values of M$_{\rm BH}$/M$_{\star}$ are coincident with a stronger suppression of SF than those galaxies in which the M$_{\rm BH}$/M$_{\star}$ value is low.

\begin{figure*}
  \centering
  \includegraphics[width=\textwidth]{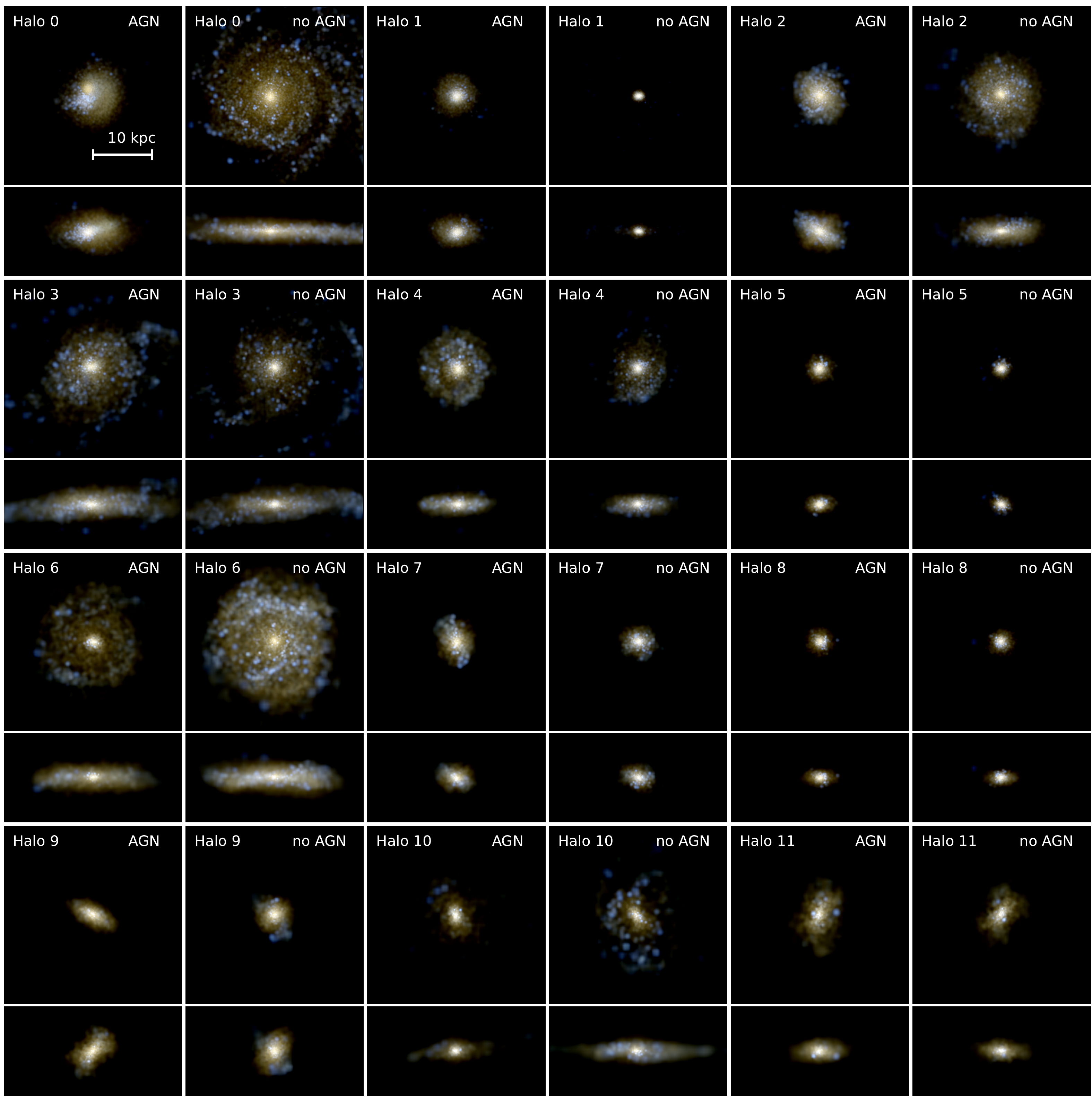}
  \caption{Face-on ($30\times 30$ kpc) and edge-on ($30\times 15$ kpc) projected stellar light images of galaxies at $z$=$0$. Each galaxy runs with the fiducial model is shown next to its no-AGN counterpart. Colours are based on the i-, g- and u-band luminosity of stars; old stars are shown in red, and young ones in blue, respectively. From top-left to bottom-right, galaxies are ordered in decreasing BH mass, as in Fig. \ref{fig:SFH}.}
  \label{fig:render}
\end{figure*}

Our results are in broad agreement with the work of \cite{Barai2019}, who find that BHs need to be at least $10^{5}$$\,\rm M_\odot$ in order to suppress SF in dwarf galaxies. Subsequent work from \citet{Sharma2020} shows that, in the \texttt{ROMULUS25} simulations, the galaxies whose star formation is most strongly suppressed are those that host over-massive BHs, while trends for the specific SFR dependence on BH mass are washed out or inverted for galaxies with  M$_{\star}$<6$\times10^9$$\,\rm M_{\odot}$ in the \texttt{FABLE} simulations \citep{Koudmani2021}. Both studies refer to cosmological simulations. Furthermore, a previous idealized simulation work \citep{Koudmani2019}  found that AGN activity in dwarfs is unlikely to regulate the global SFR of the galaxy even with an AGN shining near the Eddington luminosity. All in all, there is no current consensus on the effect of AGN feedback in reducing SF in simulated dwarf galaxies: the reasons for such discrepancies can be multiple, such as differences in resolution, BH seeding and accretion schemes, as well as details and implementations of AGN feedback models.

\subsection{Morphologies}\label{sec:morpho}

Given the findings highlighted in previous sections, we might expect to see morphological differences in dwarf galaxies run with and without AGN.\\

Fig. \ref{fig:render} shows the face-on and edge-on stellar light projections of our 12 simulated dwarf galaxies for both configurations at $z$=$0$. Each colour represents the i-, g- and u-band luminosity of the stars enclosed in a box of 30  kpc in radius. Redder and bluer colours indicate older and younger stars, respectively. In general, it can be observed that the morphology of those galaxies hosting a larger BH (top row, Halos 0-3) differ more from their counterpart than those hosting a smaller BH (bottom row, Halos 9-11).
\begin{table}
    \centering
    \begin{tabular}{lccc}
    & \multicolumn{1}{c}{AGN} & \multicolumn{1}{c}{no AGN}\\
    \cmidrule(lr){2-2}
    \cmidrule(lr){3-3}
       {Run}  & $\frac{R_{\mathrm{eff}}}{(\mathrm{kpc})}$ & $\frac{R_{\mathrm{eff}}}{(\mathrm{kpc})}$ & $\Delta \% $\\
       \midrule

      H0 &        2.83 &           3.86 &         -26 \\
      H1 &        1.90 &           1.13 &          67 \\
      H2 &        1.34 &           2.17 &         -38 \\
      H3 &        3.18 &           3.55 &         -11 \\
      H4 &        1.36 &           1.36 &          0 \\
      H5 &        0.76 &           0.61 &          25 \\
      H6 &        2.89 &           3.22 &         -10 \\
      H7 &        1.42 &           1.26 &          13    \\
      H8 &        1.06 &           1.03 &          2 \\
      H9 &        1.20 &           1.08 &          11    \\
      H10 &        2.02 &           2.29 &         -11 \\
      H11 &        1.98 &           1.91 &          3 \\

    \end{tabular}
    \caption{Effective radius  at $z$=$0$ for each simulated dwarf galaxy. In the last column, we show the percentage of change in R$_{\mathrm{eff}}$ across the AGN vs noAGN configuration.}
    \label{tab:reff}
\end{table}
However, a clear correlation between BH mass and morphological type is not immediately derivable: some galaxies appear to be more compact once AGNs are included, whilst others are more extended.
To better quantify the change in galaxy size due to AGN, the percentage of change of the effective radius between the AGN  and noAGN runs is reported in Table~\ref{tab:reff}. As anticipated, there is no evident pattern of a systematic change in dwarf galaxies' morphologies due to the presence of a centrally accreting BH.

Some previous work, based on idealized simulations of slightly larger galaxies \citep{Choi2014}, suggested that the inclusion of AGNs increases the effective radius of the simulated galaxy only when the associated feedback is mechanical. We recall that our AGN feedback model is purely thermal, which may explain the lack of a clear pattern in augmenting the R$_{\mathrm{eff}}$ of the AGN runs. Furthermore, we note that \cite{Irodotou2022} found that MW-mass \texttt{AURIGA} galaxies, in which AGN feedback was included, tend to have systematically higher effective radii than their noAGN companions:  this might indicate that thermal AGNs could be acting differently in the dwarf galaxy regime.
We conclude that the lack of a clear, systematic AGN-driven variation in morphology makes it hard to use such an observable to determine the presence/impact of AGN activity in dwarfs.

\begin{figure}
	\includegraphics[width=\columnwidth]{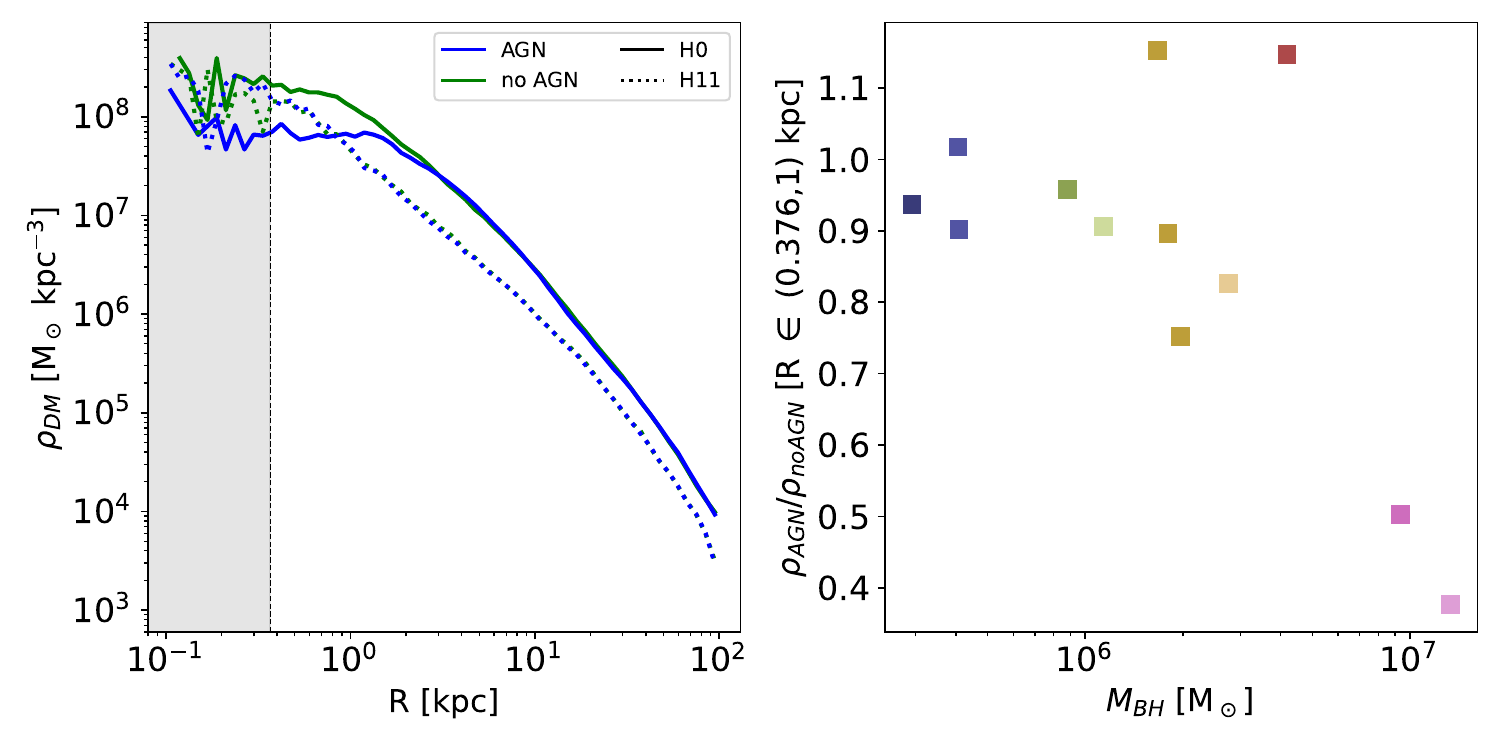}
    \caption{Left-hand panel: DM profiles for the galaxies containing the most massive (H0, solid lines) and least massive (H11, dotted lines) BHs. In green, the configuration without AGN, in blue the AGN run. The grey area shows radii below the simulations' physical softening length. Right-hand panel: Ratio between the mean DM density profile in the inner region of each galaxy (within the softening length and 1 kpc) in the AGN vs noAGN configuration, plotted against the $z$=$0$ BH mass.}
    \label{fig:DMprofiles}
\end{figure}

\subsection{DM density profiles}\label{sec:DMprof}

Theoretical studies found that AGNs may affect the  DM  mass distribution in galaxies, by flattening its inner density profile and creating DM cores \citep[]{Martizzi2012,Waterval2022}. To determine whether a similar effect occurs for our dwarf galaxies, we studied the DM profiles of our simulated sample.
We plot in Fig. \ref{fig:DMprofiles}, left panel, the DM  profile of two extreme cases: the galaxy that hosts the most massive BH, H0, and the one that contains the smallest BH, H11. AGN runs are indicated in blue, while no-AGN runs are shown in green. The fiducial halo hosting the most massive BH shows a non-negligible decrease in its inner DM density relative to its noAGN counterpart: the former has a core of about 2 kpc in size, well beyond the physical softening length of the simulation. A similar effect is observed for the galaxy hosting the second-most massive  BH (though not shown in this plot for clarity). Overall, these galaxies show up to a $\sim$65$\,\%$ reduction in their central DM  density, as opposed to the case of the least massive BH, found in  H11, which essentially does not lead to any significant change in central DM density.

\begin{figure*}
        \includegraphics[width=\textwidth]{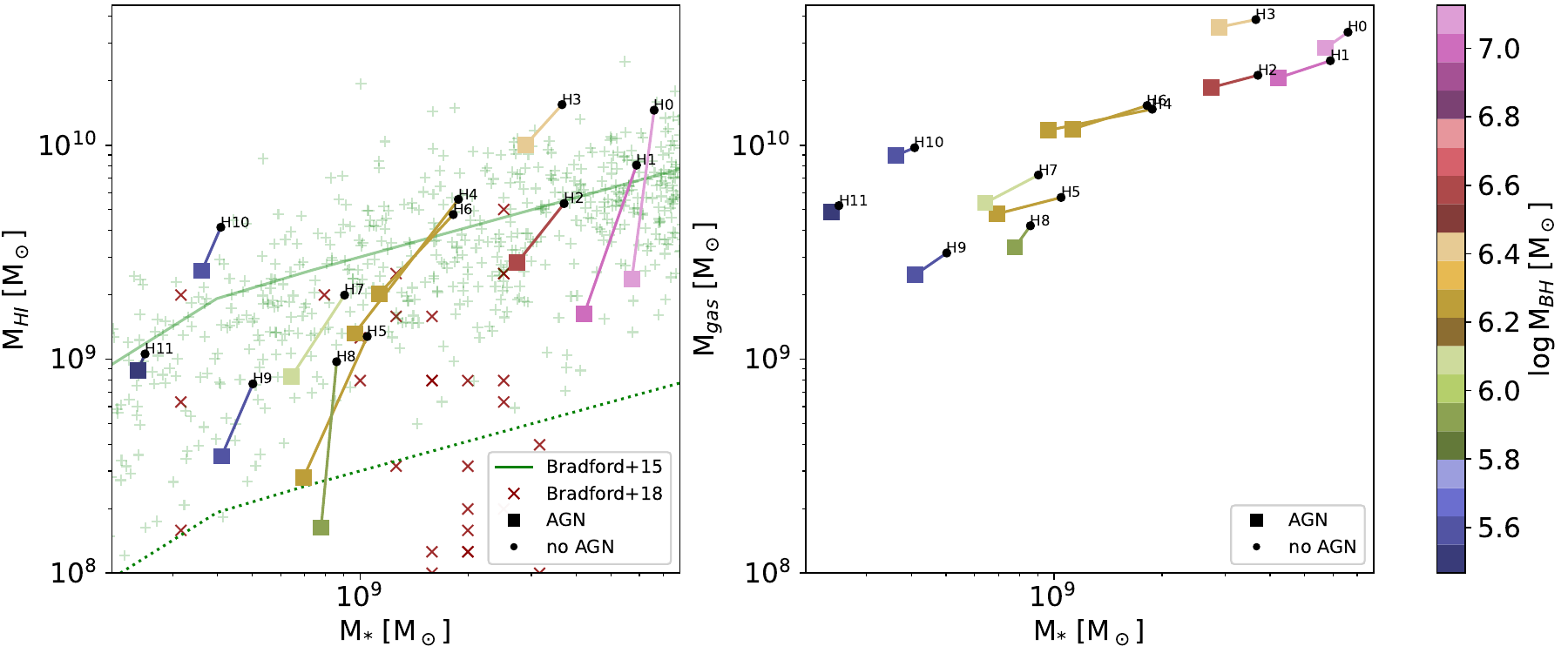}
        \caption{Left-hand panel: Neutral hydrogen mass versus stellar mass for simulated dwarfs with (coloured squares) and without (black points) AGNs. Observational data of isolated dwarfs by \protect\cite{Bradford2015} are shown as green crosses, while  AGN candidates are indicated as red crosses, from \protect\cite{Bradford2018}. The solid green line represents the best fit of the \protect\cite{Bradford2015} data, while the dotted green line is the threshold below which they define galaxies as to be quenched. Right-hand panel: total gas mass versus stellar mass for our set of simulated dwarf galaxies.}
    \label{fig:MHImasses}
\end{figure*}

We note that the galaxy harbouring the most massive BH exhibits a non-cuspy distribution irrespective of whether or not AGN feedback is included, i.e., its  DM density inner slope is already "flatter" than -1  in the simulations without AGNs (green solid line in Fig. \ref{fig:DMprofiles}, left panel). Consequently, the presence of a massive BH does not inherently imply the creation of a DM core from a cuspy profile: instead, it results in a reduction of density of whatever initial DM  distribution was already in place. In our simulations, all but the most massive galaxy (H0) are cuspy in the no-AGN run, and remain cuspy after the inclusion of the central BH. They suffer, however, a reduction in their central DM density proportional to the mass of the BH. In Fig. \ref{fig:DMprofiles}, right panel, we show the ratio between the mean of the DM density profile in the inner region of each galaxy (i.e. within radii from the softening length to 1 kpc) for the AGN and noAGN configuration, as a function of BH mass at $z$=$0$. We see minimal to no change in density for the haloes harbouring BHs less massive than 10$^6$$\,\rm M_{\odot}$, while more massive haloes produce a reduction in central DM density which is stronger and stronger as we approach the most massive BHs of our simulated dwarfs, with  M$_{\mathrm{BH}}$=$10^7$$\,\rm M_{\odot}$.

This effect is due to AGN-driven gas outflows that, similarly to what happens in the case of SNae-generated outflows, reduce the total gravitational force towards the centre of galaxies, allowing DM to move to their outskirts \citep[and references therein]{Pontzen2012,DiCintio2014}.

Previous work based on cosmological simulations of massive galaxies including BHs had studied the role of AGN in determining the final DM  mass distribution in galaxies. \citet{Maccio2020} showed that  AGN generates outflows that can partially counteract the DM baryonic contraction due to the large central stellar component in massive galaxies, effectively relaxing the central DM distribution in their simulated haloes, with masses larger than M$_{\mathrm{halo}}$=3$\times10^{12}$$\,\rm M_{\odot}$. Other simulations, including several different implementations of AGN feedback in dwarf galaxies, consistently produced cuspy DM profiles, even when the AGN worked at maximum efficiency \citep{Koudmani2022}. In such configurations, dynamic heating of DM  by AGN feedback, which would lead to the transformation of the galaxy's central region into a core, was not found. Nevertheless, a subtle yet systematic outcome is observed towards lower central densities for lower stellar-to-virial mass ratios for the AGN feedback-dominated set-ups (see Fig. 9 of \citealt{Koudmani2022}).

\subsection{Neutral hydrogen properties}\label{sec:HIproperties}

A reduced SF in those simulated dwarfs harbouring AGNs implies that such feedback affects the cold, star-forming gas in these galaxies, either by expelling it, or by heating it, or both.
In Fig. \ref{fig:MHImasses}, left-panel, we show the total amount of neutral hydrogen, HI, versus the stellar mass of each galaxy inside the virial radius. As in previous figures, the fiducial, AGN runs are shown as squares, colour-coded by the $z$=$0$ BH mass, and the noAGN ones as black points. Each AGN-noAGN galaxy pair is connected with a line, to guide the eyes.
Here, HI gas is computed following the \citet{leroy08} method as described in \citet{Marinacci2017}. 
As green crosses we show observations of isolated dwarfs from \cite{Bradford2015}, and as green solid lines their mean M$_{\mathrm{HI}}$-M$_{\star}$ relation, while as dotted-green line, we show the observational threshold below which galaxies are defined as quenched.
As red crosses, we further show isolated AGN candidates from \citet{Bradford2018}.

Fig. \ref{fig:MHImasses} clearly shows how the introduction of AGN feedback notably reduces the neutral hydrogen content of our simulated dwarf galaxies, by almost an order of magnitude in some extreme cases (halo H0, H1 and H5), approaching the quenched threshold given in \cite{Bradford2015}. Our results match the observational trends reported in \cite{Bradford2018}, in which candidate dwarfs hosting AGNs have, on average, less HI gas mass than the isolated no-AGN sample. A similar behaviour is also found in simulations by  \citet{Sharma2020} (their Fig. 12), in which galaxies hosting over-massive BHs tend to contain less neutral hydrogen, at a fixed stellar mass, than galaxies hosting less massive BHs.
In Fig. \ref{fig:MHImasses}, right panel, we show the total amount of gas versus the M$_{\star}$ of each galaxy. It can be appreciated that the reduction in HI gas between the AGN and noAGN configuration (left-hand panel) is notably larger than the corresponding reduction in total (hot and cold) gas which is, in most cases,  negligible. 
This supports the notion that AGN feedback predominantly heats the gas rather than expels it completely from the host galaxy.

\begin{figure*}
        \includegraphics[width=\textwidth]{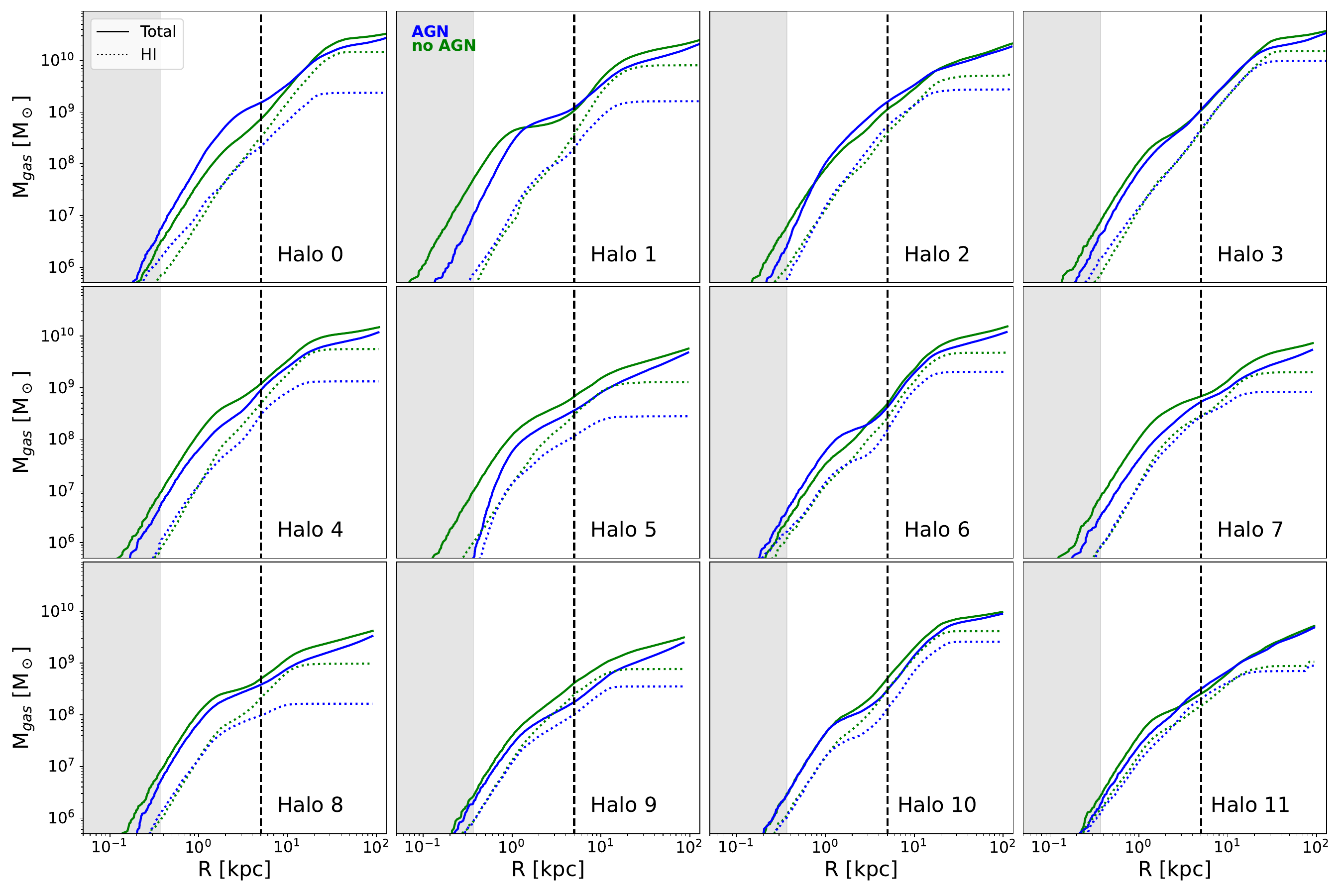}
    \caption{Cumulative gas mass vs galactic radius, for the noAGN configuration, in green, and the AGN one, in blue. Galaxies are ordered by increasing  BH mass, from top-left to bottom-right. Total gas mass is shown as solid line, while HI gas as dotted line.
 The grey area indicates  radii below the simulations' physical softening length, while the vertical line highlights a radius of 5 kpc, to facilitate comparison with the SFHs in the inner galaxy region as shown in Fig. \ref{fig:SFH}.}
    \label{fig:Mgasvsr}
\end{figure*}

Fig. \ref{fig:Mgasvsr} further shows the radial profiles of the cumulative mass of all gas (solid line) and neutral hydrogen (dotted lines) in both configurations (AGN in blue, noAGN in green), at $z$=$0$. In general, galaxies converge to a similar total gas amount at the scale of the virial radius, while this is not the case for neutral hydrogen, for which the fractional difference cumulatively builds up with radius, in some cases showing an order of magnitude more neutral hydrogen in the circumgalactic medium of the noAGN runs compared to the AGN runs (reflecting the global decrease of SF in such case).

While for the least massive galaxies (H4 to H11) there is less HI mass at 5 kpc in AGN compared to noAGN configuration, leading to a central SF suppression within this radius (visible in Fig. \ref{fig:SFH}), the four most massive haloes (H0 to H3) contain a similar amount of HI gas in AGN vs noAGN runs in their inner regions: the HI reduction happens mostly in the outskirt of the galaxies, rather than in the very central kpcs.
In these massive haloes (corresponding to M$_{\star}$$>$2$\times $10$^9$$\,\rm M_{\odot}$) AGNs are able to warm up the very central regions of galaxies at high-z, during what is typically the maximum BH accretion phase, but such gas is then able to cool back down by $z$=$0$, and indeed even form stars (see SFH of H0-H3 within 5 kpc, Fig. \ref{fig:SFH}).
We verify this claim by studying the HI mass profile vs time in the most massive galaxy, H0. At high-z, as expected, less HI was present in the AGN configuration in the inner region, compared to the noAGN one (leading to a high-z suppression of central SF in the AGN case), while over time, the HI amount in the inner 5 kpc became more similar in the AGN and noAGN case. Thus, showing an effective cooling back of such gas that leads to new episodes of SF in the centre at later times.  
In other words, in our simulations, AGNs do not seem able to \textit{keep} the gas warm in the central regions of massive dwarfs all the way to $z$=$0$, while it is able to globally reduce the HI fraction compared to galaxies run without a BH. This is also clearly seen in the SFHs of Fig. \ref{fig:SFH}: galaxies run with AGNs are not quenched in their inner few kpcs at $z$=$0$.

Existing literature shows results in partial agreement with ours. In the idealized setup of \citet{Koudmani2019}, who run a galaxy with $\rm M_{\star}$=2$\times$10$^9$ M$_{\odot}$ (therefore matching the stellar mass of our most massive haloes, H0 to H3), the authors found that in runs with AGN feedback, central SFRs are systematically
suppressed, in particular at high instantaneous star formation efficiencies $\epsilon_{\mathrm{SF}}$, while at lower $\epsilon_{\mathrm{SF}}$,  central suppression in SF is not observed by the end of their simulation (see their Fig. 6, top panels, and note the larger amount of total gas in the inner region of their galaxy in the AGN run).

Studying in detail the complex interplay between AGN feedback, outflow properties and gas angular momentum, as a function of BH and galaxy mass, is a not trivial matter that requires a dedicated analysis, tracing the history of feedback-driven outflows, inflows, and gas recycling. This is beyond the scope of this paper, and we defer such an analysis to a future work (Arjona-Gálvez et al. in prep.).

\section{Conclusions}\label{sec:Conclusions}
In recent years, a wealth of observations has revealed actively-accreting central black holes (AGNs) within dwarf galaxies. This discovery suggests a previously unexplored avenue, raising the possibility that AGNs may play a fundamental role not only in the evolution of massive galaxies but also in that of low-mass ones.
In this study, we use zoom-in cosmological simulation of 12 dwarf galaxy haloes, with $10^{8.3}$$\,\rm M_\odot$$\leq$M$_\star$$\leq$$10^{9.8}$$\,\rm M_\odot$, taken from the \texttt{AURIGA} project, to investigate the impact of AGN feedback on the evolution of such low-mass galaxies. We ran two simulations for each set of initial conditions: one with the full physics model including AGN feedback, and one without AGN feedback. The key findings of our work are summarized as follows:

\begin{itemize}

    \item The BH seeding and accretion model used generate galaxies with central BHs that reproduce the trends of well-known observational relations, such as the M$_{\rm BH}$-M$_\star$ and M$_{\rm BH}$-$\sigma_\star$ relation (Fig. \ref{fig:Mstar-Mbh} and \ref{fig:sigma-Mbh}), as well as galaxies that lie within 1 sigma scatter of the expected M$_\star$-M$_{200}$ relation (Fig. \ref{fig:SHMR}), thus representing the ideal starting point for our research;\\

    \item AGN feedback reduces the global SF of dwarf galaxies by as much as a factor of 2 for galaxies hosting BHs with masses $\gtrsim$10$^6$$\,\rm M_{\odot}$ (Fig. \ref{fig:SFH}). Most of this reduction occurs during the maximum BH accretion phase, suggesting that this type of feedback is a viable way of suppressing SF in low-mass dwarfs, even though none of our galaxies is completely quenched by $z$=$0$. The amount of SF suppression strongly correlates with the ratio between the BH and the galaxy stellar mass (Fig. \ref{fig:SFR}), indicating that the impact of AGNs  depends not only on the final BH mass but also on the global properties of the galaxy it lives in; \\

    \item No evident pattern of a systematic change in dwarf galaxies morphologies due to the presence of a centrally accreting BH is observed (Table \ref{tab:reff} and Fig. \ref{fig:render}): some galaxies are more compact once AGNs are included, while others are more extended. The dominant AGN thermal feedback used in \texttt{AURIGA} could explain the lack of a systematic change in the size (effective radius) of the galaxies;\\
  
    \item  A systematic reduction of the inner DM density in dwarfs run with AGNs is observed, proportional to the $z$=$0$ BH mass (Fig. \ref{fig:DMprofiles}): galaxies hosting BHs with $\rm M_{BH}$$\gtrsim$$10^6$$\,\rm M_\odot$ experience a significant decrease of their central DM,  up to $\sim$65$\,\%$, compared to their no-AGN counterparts, while less massive BHs lead to a negligible reduction;\\

    \item  AGN feedback is able to push gas away from the galactic centre, but it is not able to expel it completely from the galaxy virial radius: galaxies run with AGNs have a similar amount of total gas compared to galaxies run without AGNs. However, less HI is observed in the AGN configuration (Fig. \ref{fig:MHImasses}) by almost an order of magnitude in the galaxies harbouring the most massive BHs, compatible with the suppression of SF observed therein.  This in turn leads to a better match to the baryonic Tully-Fisher relation, compared to the no-AGN runs (Fig. \ref{fig:BTFR});\\
    
    \item  Studying in detail the radial profile of neutral hydrogen  (Fig. \ref{fig:Mgasvsr}), we observe a reduction of HI at all radii, including within the innermost kpc of the AGN dwarfs, compared to their noAGN companions. However, in the four most massive galaxies (M$_\star$$>$$2$$\times10^9$$\,\rm M_\odot$) HI profiles are similar in the two configurations within the central 5 kpc: such massive dwarfs have a non-negligible residual SF within their inner regions at $z$=$0$.  All in all, AGNs are able to warm up the gas and suppress star formation in the very central regions of galaxies with M$_\star$$\gtrsim$$10^9$$\,\rm M_\odot$ at high-z, during what is typically the maximum BH accretion phase, but such gas is then able to cool back down by $z$=$0$, and indeed even form stars (see SFH of H0-H3 within 5 kpc, Fig. \ref{fig:SFH}). A dedicated analysis of gas angular momentum distribution and outflow properties of dwarfs hosting AGNs is deferred to a future paper.\\

\end{itemize}

Cosmological simulations serve as a valuable tool to investigate the significance of AGN feedback in the realm of low-mass galaxies. Previous observational and theoretical studies have pinpointed the potential influence that an accreting supermassive BH can have on dwarf galaxies' evolution. Our pilot work, in which dwarf galaxies are simulated with and without the inclusion of AGNs, at high resolution and within a fully cosmological context, highlights the importance of properly modelling AGN feedback, shedding light on the impact of this scarcely studied source of feedback at the lowest mass scales.

We caution the reader that the stellar and halo mass ranges explored in this work cover the most massive dwarf galaxy regime only, and that the obtained results are dependent on the particular AGN feedback scheme implemented in the code. In the future, extending the sample to less massive dwarf galaxies will provide valuable insights into identifying the minimum BH and galaxy mass, if any, above which AGN feedback starts being important. Additionally, a comprehensive exploration employing various state-of-the-art AGN feedback schemes will be pivotal to test this hypothesis.\\

\section*{Acknowledgements}

E. Arjona-Gálvez acknowledges support from the Agencia Espacial de Investigación del Ministerio de Ciencia e Innovación (\textsc{AEI-MICIN}) and the European Social Fund (\textsc{ESF+}) through a FPI grant PRE2020-096361. A. Di Cintio is supported by a Junior Leader fellowship from ‘La Caixa’ Foundation (ID 100010434), code LCF/BQ/PR20/11770010. R. Grand acknowledges financial support from the Spanish Ministry of Science and Innovation (MICINN) through the Spanish State Research Agency, under the Severo Ochoa Program 2020-2023 (CEX2019-000920-S), and support from an STFC Ernest Rutherford Fellowship (ST/W003643/1).

This research made use of the \textit{LaPalma} HPC cluster at the Instituto de Astrofisica
de Canarias, under project \textit{can43}, PI A. Di Cintio. The authors thankfully acknowledge the technical
expertise and assistance provided by the Spanish Supercomputing
Network (Red Española de Supercomputacion, RES).

We gratefully thank Alexander Knebe and Salvador Cardona-Barrero for help with the halo-finder \texttt{AHF} and the software \texttt{PYNBODY}.


%
%

\bibliographystyle{aa}
\bibliography{ArjonaGalvez}

\begin{thebibliography}{104}
\expandafter\ifx\csname natexlab\endcsname\relax\def\natexlab#1{#1}\fi

\bibitem[{{Angl{\'e}s-Alc{\'a}zar} {et~al.}(2017){Angl{\'e}s-Alc{\'a}zar}, {Faucher-Gigu{\`e}re}, {Quataert}, {Hopkins}, {Feldmann}, {Torrey}, {Wetzel}, \& {Kere{\v{s}}}}]{Angles2017}
{Angl{\'e}s-Alc{\'a}zar}, D., {Faucher-Gigu{\`e}re}, C.-A., {Quataert}, E., {et~al.} 2017, \mnras, 472, L109

\bibitem[{{Baldassare} {et~al.}(2018){Baldassare}, {Geha}, \& {Greene}}]{Baldassare2018}
{Baldassare}, V.~F., {Geha}, M., \& {Greene}, J. 2018, \apj, 868, 152

\bibitem[{{Baldassare} {et~al.}(2020){Baldassare}, {Geha}, \& {Greene}}]{Baldassare2020}
{Baldassare}, V.~F., {Geha}, M., \& {Greene}, J. 2020, \apj, 896, 10

\bibitem[{{Baldassare} {et~al.}(2015){Baldassare}, {Reines}, {Gallo}, \& {Greene}}]{Baldassare2015}
{Baldassare}, V.~F., {Reines}, A.~E., {Gallo}, E., \& {Greene}, J.~E. 2015, \apjl, 809, L14

\bibitem[{{Baldassare} {et~al.}(2017){Baldassare}, {Reines}, {Gallo}, \& {Greene}}]{Baldassare2017}
{Baldassare}, V.~F., {Reines}, A.~E., {Gallo}, E., \& {Greene}, J.~E. 2017, \apj, 836, 20

\bibitem[{{Barai} \& {de Gouveia Dal Pino}(2019)}]{Barai2019}
{Barai}, P. \& {de Gouveia Dal Pino}, E.~M. 2019, \mnras, 487, 5549

\bibitem[{{Binney} \& {Tabor}(1995)}]{Binney1995}
{Binney}, J. \& {Tabor}, G. 1995, \mnras, 276, 663

\bibitem[{{Birchall} {et~al.}(2020){Birchall}, {Watson}, \& {Aird}}]{Birchall2020}
{Birchall}, K.~L., {Watson}, M.~G., \& {Aird}, J. 2020, \mnras, 492, 2268

\bibitem[{Bondi(1952)}]{Bondi1952}
Bondi, H. 1952, Monthly Notices of the Royal Astronomical Society, 112, 195

\bibitem[{Bondi \& Hoyle(1944)}]{Bondi1944}
Bondi, H. \& Hoyle, F. 1944, Monthly Notices of the Royal Astronomical Society, 104, 273

\bibitem[{{Boylan-Kolchin} {et~al.}(2012){Boylan-Kolchin}, {Bullock}, \& {Kaplinghat}}]{BoylanKolchin2012}
{Boylan-Kolchin}, M., {Bullock}, J.~S., \& {Kaplinghat}, M. 2012, \mnras, 422, 1203

\bibitem[{{Bradford} {et~al.}(2015){Bradford}, {Geha}, \& {Blanton}}]{Bradford2015}
{Bradford}, J.~D., {Geha}, M.~C., \& {Blanton}, M.~R. 2015, \apj, 809, 146

\bibitem[{{Bradford} {et~al.}(2018){Bradford}, {Geha}, {Greene}, {Reines}, \& {Dickey}}]{Bradford2018}
{Bradford}, J.~D., {Geha}, M.~C., {Greene}, J.~E., {Reines}, A.~E., \& {Dickey}, C.~M. 2018, \apj, 861, 50

\bibitem[{{Brook} \& {Di Cintio}(2015)}]{Brook2015}
{Brook}, C.~B. \& {Di Cintio}, A. 2015, \mnras, 450, 3920

\bibitem[{{Brook} {et~al.}(2016){Brook}, {Santos-Santos}, \& {Stinson}}]{Brook16}
{Brook}, C.~B., {Santos-Santos}, I., \& {Stinson}, G. 2016, \mnras, 459, 638

\bibitem[{{Bullock}(2010)}]{Bullock2010}
{Bullock}, J.~S. 2010, arXiv e-prints, arXiv:1009.4505

\bibitem[{{Bundy}(2015)}]{Bundy2015}
{Bundy}, K. 2015, in Galaxy Masses as Constraints of Formation Models, ed. M.~{Cappellari} \& S.~{Courteau}, Vol. 311, 100--103

\bibitem[{{Chilingarian} {et~al.}(2018){Chilingarian}, {Katkov}, {Zolotukhin}, {Grishin}, {Beletsky}, {Boutsia}, \& {Osip}}]{Chilingarian2018}
{Chilingarian}, I.~V., {Katkov}, I.~Y., {Zolotukhin}, I.~Y., {et~al.} 2018, \apj, 863, 1

\bibitem[{{Choi} {et~al.}(2014){Choi}, {Naab}, {Ostriker}, {Johansson}, \& {Moster}}]{Choi2014}
{Choi}, E., {Naab}, T., {Ostriker}, J.~P., {Johansson}, P.~H., \& {Moster}, B.~P. 2014, \mnras, 442, 440

\bibitem[{{Dashyan} {et~al.}(2018){Dashyan}, {Silk}, {Mamon}, {Dubois}, \& {Hartwig}}]{Dashyan2018}
{Dashyan}, G., {Silk}, J., {Mamon}, G.~A., {Dubois}, Y., \& {Hartwig}, T. 2018, \mnras, 473, 5698

\bibitem[{{Davis} {et~al.}(1985){Davis}, {Efstathiou}, {Frenk}, \& {White}}]{Davis1985}
{Davis}, M., {Efstathiou}, G., {Frenk}, C.~S., \& {White}, S.~D.~M. 1985, \apj, 292, 371

\bibitem[{{de Blok}(2010)}]{deBlok2010}
{de Blok}, W.~J.~G. 2010, Advances in Astronomy, 2010, 789293

\bibitem[{{Dekel} \& {Silk}(1986)}]{Dekel1986}
{Dekel}, A. \& {Silk}, J. 1986, \apj, 303, 39

\bibitem[{{Di Cintio} {et~al.}(2014){Di Cintio}, {Brook}, {Macci{\`o}}, {Stinson}, {Knebe}, {Dutton}, \& {Wadsley}}]{DiCintio2014}
{Di Cintio}, A., {Brook}, C.~B., {Macci{\`o}}, A.~V., {et~al.} 2014, \mnras, 437, 415

\bibitem[{{Di Cintio} \& {Lelli}(2016)}]{DiCintio2016}
{Di Cintio}, A. \& {Lelli}, F. 2016, Monthly Notices of the Royal Astronomical Society: Letters, 456, L127

\bibitem[{{Di Matteo} {et~al.}(2005){Di Matteo}, {Springel}, \& {Hernquist}}]{DiMatteo2005}
{Di Matteo}, T., {Springel}, V., \& {Hernquist}, L. 2005, in Growing Black Holes: Accretion in a Cosmological Context, ed. A.~{Merloni}, S.~{Nayakshin}, \& R.~A. {Sunyaev}, 340--345

\bibitem[{{Dickey} {et~al.}(2019){Dickey}, {Geha}, {Wetzel}, \& {El-Badry}}]{Dickey2019}
{Dickey}, C.~M., {Geha}, M., {Wetzel}, A., \& {El-Badry}, K. 2019, \apj, 884, 180

\bibitem[{{Dubois} {et~al.}(2014){Dubois}, {Pichon}, {Welker}, {Le Borgne}, {Devriendt}, {Laigle}, {Codis}, {Pogosyan}, {Arnouts}, {Benabed}, {Bertin}, {Blaizot}, {Bouchet}, {Cardoso}, {Colombi}, {de Lapparent}, {Desjacques}, {Gavazzi}, {Kassin}, {Kimm}, {McCracken}, {Milliard}, {Peirani}, {Prunet}, {Rouberol}, {Silk}, {Slyz}, {Sousbie}, {Teyssier}, {Tresse}, {Treyer}, {Vibert}, \& {Volonteri}}]{Dubois2014}
{Dubois}, Y., {Pichon}, C., {Welker}, C., {et~al.} 2014, \mnras, 444, 1453

\bibitem[{{Dubois} {et~al.}(2015){Dubois}, {Volonteri}, {Silk}, {Devriendt}, {Slyz}, \& {Teyssier}}]{Dubois2015}
{Dubois}, Y., {Volonteri}, M., {Silk}, J., {et~al.} 2015, \mnras, 452, 1502

\bibitem[{{Ferrarese} \& {Merritt}(2000)}]{Ferrarese2000}
{Ferrarese}, L. \& {Merritt}, D. 2000, \apjl, 539, L9

\bibitem[{{Flores} \& {Primack}(1994)}]{Flores1994}
{Flores}, R.~A. \& {Primack}, J.~R. 1994, \apjl, 427, L1

\bibitem[{{Garrison-Kimmel} {et~al.}(2014){Garrison-Kimmel}, {Boylan-Kolchin}, {Bullock}, \& {Kirby}}]{GarrisonKimmel2014}
{Garrison-Kimmel}, S., {Boylan-Kolchin}, M., {Bullock}, J.~S., \& {Kirby}, E.~N. 2014, \mnras, 444, 222

\bibitem[{{Governato} {et~al.}(2010){Governato}, {Brook}, {Mayer}, {Brooks}, {Rhee}, {Wadsley}, {Jonsson}, {Willman}, {Stinson}, {Quinn}, \& {Madau}}]{Governato2010}
{Governato}, F., {Brook}, C., {Mayer}, L., {et~al.} 2010, \nat, 463, 203

\bibitem[{{Grand} {et~al.}(2024){Grand}, {Fragkoudi}, {G{\'o}mez}, {Jenkins}, {Marinacci}, {Pakmor}, \& {Springel}}]{Grand2024}
{Grand}, R. J.~J., {Fragkoudi}, F., {G{\'o}mez}, F.~A., {et~al.} 2024, arXiv e-prints, arXiv:2401.08750

\bibitem[{Grand {et~al.}(2017)Grand, Gómez, Marinacci, Pakmor, Springel, Campbell, Frenk, Jenkins, \& White}]{Grand2017}
Grand, R. J.~J., Gómez, F.~A., Marinacci, F., {et~al.} 2017, Monthly Notices of the Royal Astronomical Society, stx071

\bibitem[{{Greene} \& {Ho}(2004)}]{Greene2004}
{Greene}, J.~E. \& {Ho}, L.~C. 2004, \apj, 610, 722

\bibitem[{{Greene} \& {Ho}(2007)}]{Greene2007}
{Greene}, J.~E. \& {Ho}, L.~C. 2007, in Astronomical Society of the Pacific Conference Series, Vol. 373, The Central Engine of Active Galactic Nuclei, ed. L.~C. {Ho} \& J.~W. {Wang}, 33

\bibitem[{{Greene} {et~al.}(2020){Greene}, {Strader}, \& {Ho}}]{Greene2020}
{Greene}, J.~E., {Strader}, J., \& {Ho}, L.~C. 2020, \araa, 58, 257

\bibitem[{{Guo} {et~al.}(2010){Guo}, {White}, {Li}, \& {Boylan-Kolchin}}]{Guo2010}
{Guo}, Q., {White}, S., {Li}, C., \& {Boylan-Kolchin}, M. 2010, \mnras, 404, 1111

\bibitem[{{Habouzit} {et~al.}(2017){Habouzit}, {Volonteri}, \& {Dubois}}]{Habouzit2017}
{Habouzit}, M., {Volonteri}, M., \& {Dubois}, Y. 2017, \mnras, 468, 3935

\bibitem[{{Haidar} {et~al.}(2022){Haidar}, {Habouzit}, {Volonteri}, {Mezcua}, {Greene}, {Neumayer}, {Angl{\'e}s-Alc{\'a}zar}, {Martin-Navarro}, {Hoyer}, {Dubois}, \& {Dav{\'e}}}]{Haidar2022}
{Haidar}, H., {Habouzit}, M., {Volonteri}, M., {et~al.} 2022, \mnras, 514, 4912

\bibitem[{{Henden} {et~al.}(2018){Henden}, {Puchwein}, {Shen}, \& {Sijacki}}]{Henden2018}
{Henden}, N.~A., {Puchwein}, E., {Shen}, S., \& {Sijacki}, D. 2018, \mnras, 479, 5385

\bibitem[{{Irodotou} {et~al.}(2022){Irodotou}, {Fragkoudi}, {Pakmor}, {Grand}, {Gadotti}, {Costa}, {Springel}, {G{\'o}mez}, \& {Marinacci}}]{Irodotou2022}
{Irodotou}, D., {Fragkoudi}, F., {Pakmor}, R., {et~al.} 2022, \mnras, 513, 3768

\bibitem[{Kauffmann {et~al.}(1993)Kauffmann, White, \& Guiderdoni}]{Kauffmann1993}
Kauffmann, G., White, S. D.~M., \& Guiderdoni, B. 1993, Monthly Notices of the Royal Astronomical Society, 264, 201

\bibitem[{Klypin {et~al.}(1999)Klypin, Kravtsov, Valenzuela, \& Prada}]{Klypin1999}
Klypin, A., Kravtsov, A.~V., Valenzuela, O., \& Prada, F. 1999, The Astrophysical Journal, 522, 82

\bibitem[{Knollmann \& Knebe(2009)}]{Knollmann2009}
Knollmann, S.~R. \& Knebe, A. 2009, The Astrophysical Journal Supplement Series, 182, 608

\bibitem[{{Koudmani} {et~al.}(2021){Koudmani}, {Henden}, \& {Sijacki}}]{Koudmani2021}
{Koudmani}, S., {Henden}, N.~A., \& {Sijacki}, D. 2021, \mnras, 503, 3568

\bibitem[{{Koudmani} {et~al.}(2019){Koudmani}, {Sijacki}, {Bourne}, \& {Smith}}]{Koudmani2019}
{Koudmani}, S., {Sijacki}, D., {Bourne}, M.~A., \& {Smith}, M.~C. 2019, \mnras, 484, 2047

\bibitem[{{Koudmani} {et~al.}(2022){Koudmani}, {Sijacki}, \& {Smith}}]{Koudmani2022}
{Koudmani}, S., {Sijacki}, D., \& {Smith}, M.~C. 2022, \mnras, 516, 2112

\bibitem[{{Larson}(1974)}]{Larson1974}
{Larson}, R.~B. 1974, \mnras, 169, 229

\bibitem[{{Leroy} {et~al.}(2008){Leroy}, {Walter}, {Brinks}, {Bigiel}, {de Blok}, {Madore}, \& {Thornley}}]{leroy08}
{Leroy}, A.~K., {Walter}, F., {Brinks}, E., {et~al.} 2008, \aj, 136, 2782

\bibitem[{{Macci{\`o}} {et~al.}(2020){Macci{\`o}}, {Crespi}, {Blank}, \& {Kang}}]{Maccio2020}
{Macci{\`o}}, A.~V., {Crespi}, S., {Blank}, M., \& {Kang}, X. 2020, \mnras, 495, L46

\bibitem[{{Marinacci} {et~al.}(2017){Marinacci}, {Grand}, {Pakmor}, {Springel}, {G{\'o}mez}, {Frenk}, \& {White}}]{Marinacci2017}
{Marinacci}, F., {Grand}, R. J.~J., {Pakmor}, R., {et~al.} 2017, \mnras, 466, 3859

\bibitem[{{Mart{\'\i}n-Navarro} \& {Mezcua}(2018)}]{MartinNavarro2018}
{Mart{\'\i}n-Navarro}, I. \& {Mezcua}, M. 2018, \apjl, 855, L20

\bibitem[{Martizzi {et~al.}(2012)Martizzi, Teyssier, Moore, \& Wentz}]{Martizzi2012}
Martizzi, D., Teyssier, R., Moore, B., \& Wentz, T. 2012, Monthly Notices of the Royal Astronomical Society, 422, 3081

\bibitem[{McGaugh(2012)}]{McGaugh2012}
McGaugh, S.~S. 2012, The Astronomical Journal, 143, 40

\bibitem[{{Menon} {et~al.}(2015){Menon}, {Wesolowski}, {Zheng}, {Jetley}, {Kale}, {Quinn}, \& {Governato}}]{Menon2015}
{Menon}, H., {Wesolowski}, L., {Zheng}, G., {et~al.} 2015, Computational Astrophysics and Cosmology, 2, 1

\bibitem[{{Mezcua}(2017)}]{Mezcua2017}
{Mezcua}, M. 2017, International Journal of Modern Physics D, 26, 1730021

\bibitem[{{Mezcua} {et~al.}(2016){Mezcua}, {Civano}, {Fabbiano}, {Miyaji}, \& {Marchesi}}]{Mezcua2016}
{Mezcua}, M., {Civano}, F., {Fabbiano}, G., {Miyaji}, T., \& {Marchesi}, S. 2016, \apj, 817, 20

\bibitem[{{Mezcua} {et~al.}(2018{\natexlab{a}}){Mezcua}, {Civano}, {Marchesi}, {Suh}, {Fabbiano}, \& {Volonteri}}]{Mezcua2018a}
{Mezcua}, M., {Civano}, F., {Marchesi}, S., {et~al.} 2018{\natexlab{a}}, \mnras, 478, 2576

\bibitem[{{Mezcua} \& {Dom{\'\i}nguez S{\'a}nchez}(2020)}]{Mezcua2020}
{Mezcua}, M. \& {Dom{\'\i}nguez S{\'a}nchez}, H. 2020, \apjl, 898, L30

\bibitem[{{Mezcua} {et~al.}(2018{\natexlab{b}}){Mezcua}, {Kim}, {Ho}, \& {Lonsdale}}]{Mezcua2018b}
{Mezcua}, M., {Kim}, M., {Ho}, L.~C., \& {Lonsdale}, C.~J. 2018{\natexlab{b}}, \mnras, 480, L74

\bibitem[{{Mezcua} {et~al.}(2019){Mezcua}, {Suh}, \& {Civano}}]{Mezcua2019}
{Mezcua}, M., {Suh}, H., \& {Civano}, F. 2019, \mnras, 488, 685

\bibitem[{{Moore}(1994)}]{Moore1994}
{Moore}, B. 1994, \nat, 370, 629

\bibitem[{{Moore} {et~al.}(1999){Moore}, {Ghigna}, {Governato}, {Lake}, {Quinn}, {Stadel}, \& {Tozzi}}]{Moore1999a}
{Moore}, B., {Ghigna}, S., {Governato}, F., {et~al.} 1999, \apjl, 524, L19

\bibitem[{{Moster} {et~al.}(2013){Moster}, {Naab}, \& {White}}]{Moster2013}
{Moster}, B.~P., {Naab}, T., \& {White}, S. D.~M. 2013, \mnras, 428, 3121

\bibitem[{Nulsen \& Fabian(2000)}]{Nulsen2000}
Nulsen, P. E.~J. \& Fabian, A.~C. 2000, Monthly Notices of the Royal Astronomical Society, 311, 346

\bibitem[{{Okamoto} {et~al.}(2008){Okamoto}, {Gao}, \& {Theuns}}]{Okamoto2008}
{Okamoto}, T., {Gao}, L., \& {Theuns}, T. 2008, \mnras, 390, 920

\bibitem[{{Oman} {et~al.}(2015){Oman}, {Navarro}, {Fattahi}, {Frenk}, {Sawala}, {White}, {Bower}, {Crain}, {Furlong}, {Schaller}, {Schaye}, \& {Theuns}}]{Oman2015}
{Oman}, K.~A., {Navarro}, J.~F., {Fattahi}, A., {et~al.} 2015, \mnras, 452, 3650

\bibitem[{{Pakmor} {et~al.}(2017){Pakmor}, {G{\'o}mez}, {Grand}, {Marinacci}, {Simpson}, {Springel}, {Campbell}, {Frenk}, {Guillet}, {Pfrommer}, \& {White}}]{PGG17}
{Pakmor}, R., {G{\'o}mez}, F.~A., {Grand}, R.~J.~J., {et~al.} 2017, \mnras, 469, 3185

\bibitem[{{Pakmor} {et~al.}(2016){Pakmor}, {Springel}, {Bauer}, {Mocz}, {Munoz}, {Ohlmann}, {Schaal}, \& {Zhu}}]{Pakmor15}
{Pakmor}, R., {Springel}, V., {Bauer}, A., {et~al.} 2016, \mnras, 455, 1134

\bibitem[{{Pawlik} \& {Schaye}(2009)}]{Pawlik2009}
{Pawlik}, A.~H. \& {Schaye}, J. 2009, \mnras, 396, L46

\bibitem[{{Penny} {et~al.}(2018){Penny}, {Masters}, {Smethurst}, {Nichol}, {Krawczyk}, {Bizyaev}, {Greene}, {Liu}, {Marinelli}, {Rembold}, {Riffel}, {Ilha}, {Wylezalek}, {Andrews}, {Bundy}, {Drory}, {Oravetz}, \& {Pan}}]{Penny2018}
{Penny}, S.~J., {Masters}, K.~L., {Smethurst}, R., {et~al.} 2018, \mnras, 476, 979

\bibitem[{{Pillepich} {et~al.}(2018){Pillepich}, {Springel}, {Nelson}, {Genel}, {Naiman}, {Pakmor}, {Hernquist}, {Torrey}, {Vogelsberger}, {Weinberger}, \& {Marinacci}}]{Annalisa2018}
{Pillepich}, A., {Springel}, V., {Nelson}, D., {et~al.} 2018, \mnras, 473, 4077

\bibitem[{{Planck Collaboration} {et~al.}(2014){Planck Collaboration}, {Ade}, {Aghanim}, {Armitage-Caplan}, {Arnaud}, {Ashdown}, {Atrio-Barandela}, {Aumont}, {Baccigalupi}, {Banday}, {Barreiro}, {Bartlett}, {Battaner}, {Benabed}, {Beno{\^\i}t}, {Benoit-L{\'e}vy}, {Bernard}, {Bersanelli}, {Bielewicz}, {Bobin}, {Bock}, {Bonaldi}, {Bond}, {Borrill}, {Bouchet}, {Bridges}, {Bucher}, {Burigana}, {Butler}, {Calabrese}, {Cappellini}, {Cardoso}, {Catalano}, {Challinor}, {Chamballu}, {Chary}, {Chen}, {Chiang}, {Chiang}, {Christensen}, {Church}, {Clements}, {Colombi}, {Colombo}, {Couchot}, {Coulais}, {Crill}, {Curto}, {Cuttaia}, {Danese}, {Davies}, {Davis}, {de Bernardis}, {de Rosa}, {de Zotti}, {Delabrouille}, {Delouis}, {D{\'e}sert}, {Dickinson}, {Diego}, {Dolag}, {Dole}, {Donzelli}, {Dor{\'e}}, {Douspis}, {Dunkley}, {Dupac}, {Efstathiou}, {Elsner}, {En{\ss}lin}, {Eriksen}, {Finelli}, {Forni}, {Frailis}, {Fraisse}, {Franceschi}, {Gaier}, {Galeotta}, {Galli}, {Ganga}, {Giard}, {Giardino}, {Giraud-H{\'e}raud},
  {Gjerl{\o}w}, {Gonz{\'a}lez-Nuevo}, {G{\'o}rski}, {Gratton}, {Gregorio}, {Gruppuso}, {Gudmundsson}, {Haissinski}, {Hamann}, {Hansen}, {Hanson}, {Harrison}, {Henrot-Versill{\'e}}, {Hern{\'a}ndez-Monteagudo}, {Herranz}, {Hildebrandt}, {Hivon}, {Hobson}, {Holmes}, {Hornstrup}, {Hou}, {Hovest}, {Huffenberger}, {Jaffe}, {Jaffe}, {Jewell}, {Jones}, {Juvela}, {Keih{\"a}nen}, {Keskitalo}, {Kisner}, {Kneissl}, {Knoche}, {Knox}, {Kunz}, {Kurki-Suonio}, {Lagache}, {L{\"a}hteenm{\"a}ki}, {Lamarre}, {Lasenby}, {Lattanzi}, {Laureijs}, {Lawrence}, {Leach}, {Leahy}, {Leonardi}, {Le{\'o}n-Tavares}, {Lesgourgues}, {Lewis}, {Liguori}, {Lilje}, {Linden-V{\o}rnle}, {L{\'o}pez-Caniego}, {Lubin}, {Mac{\'\i}as-P{\'e}rez}, {Maffei}, {Maino}, {Mandolesi}, {Maris}, {Marshall}, {Martin}, {Mart{\'\i}nez-Gonz{\'a}lez}, {Masi}, {Massardi}, {Matarrese}, {Matthai}, {Mazzotta}, {Meinhold}, {Melchiorri}, {Melin}, {Mendes}, {Menegoni}, {Mennella}, {Migliaccio}, {Millea}, {Mitra}, {Miville-Desch{\^e}nes}, {Moneti}, {Montier}, {Morgante},
  {Mortlock}, {Moss}, {Munshi}, {Murphy}, {Naselsky}, {Nati}, {Natoli}, {Netterfield}, {N{\o}rgaard-Nielsen}, {Noviello}, {Novikov}, {Novikov}, {O'Dwyer}, {Osborne}, {Oxborrow}, {Paci}, {Pagano}, {Pajot}, {Paladini}, {Paoletti}, {Partridge}, {Pasian}, {Patanchon}, {Pearson}, {Pearson}, {Peiris}, {Perdereau}, {Perotto}, {Perrotta}, {Pettorino}, {Piacentini}, {Piat}, {Pierpaoli}, {Pietrobon}, {Plaszczynski}, {Platania}, {Pointecouteau}, {Polenta}, {Ponthieu}, {Popa}, {Poutanen}, {Pratt}, {Pr{\'e}zeau}, {Prunet}, {Puget}, {Rachen}, {Reach}, {Rebolo}, {Reinecke}, {Remazeilles}, {Renault}, {Ricciardi}, {Riller}, {Ristorcelli}, {Rocha}, {Rosset}, {Roudier}, {Rowan-Robinson}, {Rubi{\~n}o-Mart{\'\i}n}, {Rusholme}, {Sandri}, {Santos}, {Savelainen}, {Savini}, {Scott}, {Seiffert}, {Shellard}, {Spencer}, {Starck}, {Stolyarov}, {Stompor}, {Sudiwala}, {Sunyaev}, {Sureau}, {Sutton}, {Suur-Uski}, {Sygnet}, {Tauber}, {Tavagnacco}, {Terenzi}, {Toffolatti}, {Tomasi}, {Tristram}, {Tucci}, {Tuovinen}, {T{\"u}rler}, {Umana},
  {Valenziano}, {Valiviita}, {Van Tent}, {Vielva}, {Villa}, {Vittorio}, {Wade}, {Wandelt}, {Wehus}, {White}, {White}, {Wilkinson}, {Yvon}, {Zacchei}, \& {Zonca}}]{Planck2014}
{Planck Collaboration}, {Ade}, P.~A.~R., {Aghanim}, N., {et~al.} 2014, \aap, 571, A16

\bibitem[{{Pontzen} \& {Governato}(2012)}]{Pontzen2012}
{Pontzen}, A. \& {Governato}, F. 2012, \mnras, 421, 3464

\bibitem[{{Pontzen} {et~al.}(2013){Pontzen}, {Ro{\v{s}}kar}, {Stinson}, \& {Woods}}]{Pontzen2013}
{Pontzen}, A., {Ro{\v{s}}kar}, R., {Stinson}, G., \& {Woods}, R. 2013, {pynbody: N-Body/SPH analysis for python}, Astrophysics Source Code Library, record ascl:1305.002

\bibitem[{{Read} {et~al.}(2006){Read}, {Wilkinson}, {Evans}, {Gilmore}, \& {Kleyna}}]{Read2006}
{Read}, J.~I., {Wilkinson}, M.~I., {Evans}, N.~W., {Gilmore}, G., \& {Kleyna}, J.~T. 2006, \mnras, 367, 387

\bibitem[{{Reines} {et~al.}(2013){Reines}, {Greene}, \& {Geha}}]{Reines2013}
{Reines}, A.~E., {Greene}, J.~E., \& {Geha}, M. 2013, \apj, 775, 116

\bibitem[{{Reines} \& {Volonteri}(2015)}]{Reines2015}
{Reines}, A.~E. \& {Volonteri}, M. 2015, \apj, 813, 82

\bibitem[{{Sales} {et~al.}(2022){Sales}, {Wetzel}, \& {Fattahi}}]{Sales2022}
{Sales}, L.~V., {Wetzel}, A., \& {Fattahi}, A. 2022, Nature Astronomy, 6, 897

\bibitem[{{Santos-Santos} {et~al.}(2018){Santos-Santos}, {Di Cintio}, {Brook}, {Macci{\`o}}, {Dutton}, \& {Dom{\'\i}nguez-Tenreiro}}]{SantosSantos2018}
{Santos-Santos}, I.~M., {Di Cintio}, A., {Brook}, C.~B., {et~al.} 2018, \mnras, 473, 4392

\bibitem[{{Santos-Santos} {et~al.}(2020){Santos-Santos}, {Navarro}, {Robertson}, {Ben{\'\i}tez-Llambay}, {Oman}, {Lovell}, {Frenk}, {Ludlow}, {Fattahi}, \& {Ritz}}]{SantosSantos2020}
{Santos-Santos}, I. M.~E., {Navarro}, J.~F., {Robertson}, A., {et~al.} 2020, \mnras, 495, 58

\bibitem[{{Sawala} {et~al.}(2016){Sawala}, {Frenk}, {Fattahi}, {Navarro}, {Bower}, {Crain}, {Dalla Vecchia}, {Furlong}, {Helly}, {Jenkins}, {Oman}, {Schaller}, {Schaye}, {Theuns}, {Trayford}, \& {White}}]{Sawala2016}
{Sawala}, T., {Frenk}, C.~S., {Fattahi}, A., {et~al.} 2016, \mnras, 457, 1931

\bibitem[{{Schaye} {et~al.}(2015){Schaye}, {Crain}, {Bower}, {Furlong}, {Schaller}, {Theuns}, {Dalla Vecchia}, {Frenk}, {McCarthy}, {Helly}, {Jenkins}, {Rosas-Guevara}, {White}, {Baes}, {Booth}, {Camps}, {Navarro}, {Qu}, {Rahmati}, {Sawala}, {Thomas}, \& {Trayford}}]{Schaye2015}
{Schaye}, J., {Crain}, R.~A., {Bower}, R.~G., {et~al.} 2015, \mnras, 446, 521

\bibitem[{{Schramm} \& {Silverman}(2013)}]{Schramm2013}
{Schramm}, M. \& {Silverman}, J.~D. 2013, \apj, 767, 13

\bibitem[{{Sharma} {et~al.}(2020){Sharma}, {Brooks}, {Somerville}, {Tremmel}, {Bellovary}, {Wright}, \& {Quinn}}]{Sharma2020}
{Sharma}, R.~S., {Brooks}, A.~M., {Somerville}, R.~S., {et~al.} 2020, \apj, 897, 103

\bibitem[{{Shen} {et~al.}(2010){Shen}, {Wadsley}, \& {Stinson}}]{Shen2010}
{Shen}, S., {Wadsley}, J., \& {Stinson}, G. 2010, \mnras, 407, 1581

\bibitem[{Silk(2017)}]{Silk2017}
Silk, J. 2017, The Astrophysical Journal, 839, L13

\bibitem[{{Somerville}(2002)}]{Somerville2002}
{Somerville}, R.~S. 2002, \apjl, 572, L23

\bibitem[{Springel(2005)}]{Springel2005}
Springel, V. 2005, The cosmological simulation code GADGET-2

\bibitem[{Springel(2010)}]{Springel2010}
Springel, V. 2010, Monthly Notices of the Royal Astronomical Society, 401, 791

\bibitem[{{Springel} {et~al.}(2005){Springel}, {Di Matteo}, \& {Hernquist}}]{Springel2005b}
{Springel}, V., {Di Matteo}, T., \& {Hernquist}, L. 2005, \mnras, 361, 776

\bibitem[{{Springel} \& {Hernquist}(2003)}]{SH03}
{Springel}, V. \& {Hernquist}, L. 2003, \mnras, 339, 289

\bibitem[{Stark {et~al.}(2009)Stark, McGaugh, \& Swaters}]{Stark2009}
Stark, D.~V., McGaugh, S.~S., \& Swaters, R.~A. 2009, The Astronomical Journal, 138, 392

\bibitem[{{Stinson} {et~al.}(2006){Stinson}, {Seth}, {Katz}, {Wadsley}, {Governato}, \& {Quinn}}]{Stinson2006}
{Stinson}, G., {Seth}, A., {Katz}, N., {et~al.} 2006, \mnras, 373, 1074

\bibitem[{{Trebitsch} {et~al.}(2018){Trebitsch}, {Volonteri}, {Dubois}, \& {Madau}}]{Trebitsch2018}
{Trebitsch}, M., {Volonteri}, M., {Dubois}, Y., \& {Madau}, P. 2018, \mnras, 478, 5607

\bibitem[{{Vogelsberger} {et~al.}(2013){Vogelsberger}, {Genel}, {Sijacki}, {Torrey}, {Springel}, \& {Hernquist}}]{Vogelsberger2013}
{Vogelsberger}, M., {Genel}, S., {Sijacki}, D., {et~al.} 2013, \mnras, 436, 3031

\bibitem[{{Vogelsberger} {et~al.}(2014){Vogelsberger}, {Genel}, {Springel}, {Torrey}, {Sijacki}, {Xu}, {Snyder}, {Nelson}, \& {Hernquist}}]{Vogelsberger2014a}
{Vogelsberger}, M., {Genel}, S., {Springel}, V., {et~al.} 2014, \mnras, 444, 1518

\bibitem[{{Wadsley} {et~al.}(2017){Wadsley}, {Keller}, \& {Quinn}}]{Wadsley2017}
{Wadsley}, J.~W., {Keller}, B.~W., \& {Quinn}, T.~R. 2017, \mnras, 471, 2357

\bibitem[{Waterval {et~al.}(2022)Waterval, Elgamal, Nori, Pasquato, Macciò, Blank, Dixon, Kang, \& Ibrayev}]{Waterval2022}
Waterval, S., Elgamal, S., Nori, M., {et~al.} 2022

\bibitem[{{Weinberger} {et~al.}(2017){Weinberger}, {Springel}, {Hernquist}, {Pillepich}, {Marinacci}, {Pakmor}, {Nelson}, {Genel}, {Vogelsberger}, {Naiman}, \& {Torrey}}]{Weinberger2017}
{Weinberger}, R., {Springel}, V., {Hernquist}, L., {et~al.} 2017, \mnras, 465, 3291

\bibitem[{{White} \& {Frenk}(1991)}]{White1991}
{White}, S. D.~M. \& {Frenk}, C.~S. 1991, \apj, 379, 52

\bibitem[{{Xiao} {et~al.}(2011){Xiao}, {Barth}, {Greene}, {Ho}, {Bentz}, {Ludwig}, \& {Jiang}}]{Xiao2011}
{Xiao}, T., {Barth}, A.~J., {Greene}, J.~E., {et~al.} 2011, \apj, 739, 28

\end{thebibliography}

\end{document}